\documentclass[12pt, letterpaper]{article}

 \usepackage{setspace}
\usepackage{moreverb}
\usepackage{amsmath, amsthm, amssymb,bm,bbm}
\usepackage{tikz}
\usetikzlibrary{positioning,shapes.geometric,graphs, arrows.meta}
\usetikzlibrary[graphs]
\usepackage{color}
\usepackage{multirow}
\usepackage{lscape} 
\usepackage{geometry}
\usepackage{caption}
\usepackage[export]{adjustbox}
\usepackage{hyperref}
\usepackage{float}
\usepackage{comment}
\usepackage{multicol}
\usepackage{lipsum}
\usepackage{graphicx,subcaption} 

\usepackage{graphicx}
\usepackage{enumerate}
\usepackage[authoryear]{natbib}
\usepackage{url}  
\usepackage[english]{babel}
 \usepackage{color}
\usepackage{multirow}
\usepackage{lscape}
\usepackage{csquotes}
\usepackage{verbatim}
 \usepackage{prodint}
\usepackage[english]{babel}
\usepackage[utf8]{inputenc}
\usepackage{amsmath, amsthm, amssymb,amsfonts,bm}
\usepackage{tikz}
\usepackage{courier}
\usetikzlibrary{positioning,shapes.geometric,graphs, arrows.meta}
\usetikzlibrary[graphs]
\usepackage{color}
\usepackage{verbatim}
\usepackage{hyperref}
\usepackage{float}
\usepackage{comment}
\usepackage{mathrsfs}
\usepackage{multicol}
\usepackage{bbm}
\usepackage{lipsum}
  \usepackage{color}
\usepackage{graphicx,subcaption}  
\usepackage{float}
\usepackage{comment}
\usepackage{mathrsfs}
 \usepackage{color}
 
  \RequirePackage{amsthm,amsmath,amsfonts,amssymb}

\RequirePackage[authoryear]{natbib}
\usepackage[authoryear]{natbib}

\usepackage{geometry}

\begin{document}

\title{Estimating the Marginal Effect of a Continuous Exposure on an Ordinal Outcome using Data Subject to Covariate-Driven Treatment and Visit Processes}

\author{Janie Coulombe, Erica E. M. Moodie, and Robert W. Platt}
\date{}
\maketitle

\begin{abstract}
 In the statistical literature, a number of methods have been proposed to ensure valid inference about marginal effects of variables on a longitudinal outcome in settings with irregular monitoring times. However, the potential  biases due to covariate-driven monitoring times and confounding have rarely been considered simultaneously, and never in a setting with an ordinal outcome and a continuous exposure. In this work, we propose and demonstrate a methodology for causal inference in such a setting,  relying on a proportional odds model to study the effect of the exposure on the outcome. Irregular observation times are considered via a proportional rate model, and a generalization of inverse probability of treatment weights is used to account for the continuous exposure. We motivate our methodology by the estimation of the marginal (causal) effect of the time spent on video or computer games on suicide attempts in the \textit{Add Health} study, a longitudinal study in the United States. Although in the \textit{Add Health} data, observation times are pre-specified, our proposed approach is applicable even in more general settings such as when analyzing data from electronic health records where observations are highly irregular. In simulation studies, we let observation times vary across individuals and demonstrate that not accounting for biasing imbalances due to the monitoring and the exposure schemes can bias the estimate for the marginal odds ratio of exposure.
  
  \end{abstract}
 
\section{Introduction}

Suppose a setting where longitudinal data are assessed irregularly across patients. In particular, we focus on non-experimental settings where, contrary to randomized studies, the mere fact of being exposed to a certain level of an exposure can be related to individuals' characteristics. Our interest lies in the estimation of the causal effect of a continuous exposure on a longitudinal, ordinal (categorical) outcome which is measured irregularly. Assuming that the observation of the outcome process coincides with monitoring times, which depend on patients' characteristics, we can then  think of the outcome process as being missing  \textit{ at random} over the course of follow-up time. In observational data, imbalances such as those due to covariate-driven monitoring times or confounding can bias the estimators for the marginal effect of variables (e.g.~that of an exposure). The bias due to covariate-driven monitoring times relates to \textit{selection bias}, often due to a phenomenon called \textit{collider-stratification bias} \citep{greenland2003quantifying} where restricting a study cohort on a covariate, or conditioning on a selection indicator (which both could be colliders on the path from the exposure to the outcome) may induce dependence between the exposure and outcome. Selection bias is most often discussed in a cross-sectional setting, where it refers to  differences in covariates between the individuals selected into the study and those who were not selected. In a longitudinal setting, these imbalances related to selection can be repeated all throughout follow-up time, due to irregular visit times. For causal inference on an exposure effect, monitoring times may have to be modelled continuously in time if they are related to covariates that are simultaneously associated with the study outcome and the study exposure. 

Covariate-driven monitoring times can arise both in a study with presspecified observation times or when using data from electronic health records. We propose and demonstrate a methodology for the estimation of the causal effect of a continuous exposure on a categorical outcome that applies in both settings where observation times are pre-specified (such as the \textit{Add Health} study) or are irregular across individuals (such as when using data from electronic health records). The methodology allows for observation times  to depend on individuals' characteristics, and also accounts for potential confounders of the relationship under study. We model the monitoring indicators (which indicate whether there is a visit, or not, during which the outcome is assessed) using a proportional intensity model and the exposure model using a generalized propensity score which is then used to adjust for confounding via inverse probability of treatment weights. The outcome is modelled as a function of the exposure, using a proportional odds model.  

Our aim, to estimate the marginal effect of the time spent on video games (weekly) on suicide attempts, motivates the methodological developments. We conduct our analysis  using  longitudinal data from the first four waves of the \textit{Add Health} study in the United States \citep{harris1}. In that study, individuals were followed from their adolescence and until adulthood, and were assessed at different points in time. Individuals' personal data were collected via several questionnaires, and the collected information included demographics, social, biological and behavioral factors as well as parental factors. %, and more.   

This manuscript is divided as follows: in Section 2, we discuss the background, our assumptions and notation, as well as the proposed methodology for estimating the causal effect of a continuous exposure on a categorical, ordinal outcome. In Section 3, we present the details of the simulation studies that were conducted and corresponding results. The application of the proposed methodology to data from the \textit{Add Health} study to estimate the effect of the number of hours spent on video games on suicide attempts is presented in Section 4. A discussion follows in Section 5.
 
\section{Methods}

\subsection{Background}

This work extends previous research \citep{coulombe} in which two estimators were proposed for the marginal effect of a binary (time-varying) exposure $\mathbf{I(t)}$ on a longitudinal continuous outcome $\mathbf{Y(t)}$. These estimators were also proposed for settings where data are subject to being recorded (or monitored) at covariate-dependent visit times, allowed to be irregular across patients, and where there is potential confounding. Inverse weights were used to adjust both for confounding and bias due to informative monitoring times. In addition, the monitoring weights were allowed to be functions of mediators of the relationship of interest. %, which is convenient in the causal inference setting when we aim for an estimate of the marginal effect of exposure.

In that previous work, two estimators were proposed and compared, namely:
\begin{itemize}
\item a first estimator which extended the estimating equations proposed by \cite{lin2001semiparametric} to account for confounding via a generalized inverse probability of treatment weight, and incorporated the stabilized weight proposed by \cite{buuvzkova2009semiparametric} for the visit intensity (denoted by $\hat{\beta}_{IPCTM}$ where IPCTM denotes the inverse probability of centered treatment and monitoring estimator); and
\item a second estimator which used the standard weighted least squares estimating equations, weighted with both an inverse probability of treatment weight and an inverse intensity of visit weight. That estimator also included a cubic spline basis as a function of time since cohort entry to allow for more flexibility in modelling the effect of time since study baseline on the mean outcome. It was denoted by $\hat{\beta}_{FIPTM}$, where FIPTM refers to the flexible inverse probability of treatment and monitoring estimator.
\end{itemize}
Both estimators were demonstrated via theoretical derivations to be unbiased for large samples, and extensive simulation studies showed that both were unbiased in finite samples and their variances were quite comparable. However, $\hat{\beta}_{FIPTM}$ was easier to implement. To answer the current research question, we aim to extend this estimator to the setting where exposure is continuous, and where the outcome is categorical and ordinal. For that, we use the proportional odds model.

\subsubsection{Proportional Odds Model}
Suppose that one is interested in the association between a set of covariates $ \mathbf{X(t)}$ (which could contain the exposure of interest) and the categorical outcome $\mathbf{Y(t)}$, and let $i$ be a patient index, and $t$ denote time. Then, the proportional odds model (POM) proposed by \cite{mccullagh1980regression} models the outcome cumulative probabilities as
\begin{align}
P(Y_i(t) \le j |  \mathbf{X_i(t)} ) %= \frac{\exp(\alpha_j - \bm{\beta}' \mathbf{X_i(t)})}{1+\exp(\alpha_j - \bm{\beta}' \mathbf{X_i(t)})} \nonumber \\
= \text{expit}(\alpha_j - \bm{\beta}' \mathbf{X_i(t)}) \label{eqintro}
\end{align}
$\forall j=1,...,J$, where the expit function is the inverse logit function and the coefficients $\bm{\alpha}$ are category-specific intercepts. The effects of covariates $\mathbf{X(t)}$ are assumed to be constant for all $j$ (that is, $\bm{\beta}_j=\bm{\beta}$). Effectively, the POM is an extension of the logistic regression model to the multinomial case where, rather than merely comparing the probability of an event occurring (as compared to no event), the outcome domain is stratified at each possible ``split value'' between two subsequent ordered categories $j$ and $j+1$ in $\left\{ 1,...,J\right\}$. The probability that is modelled is thus that for the outcome to be smaller or equal to some value in the outcome domain, as opposed to larger. 

When studying the marginal effect of a variable $\mathbf{D(t)} \subset \mathbf{X(t)}$ (suppose, the exposure) on the ordinal outcome $\mathbf{Y(t)}$, different quantites may be of interest. If the outcome is made of only two categories, the POM is equivalent to the logistic regression model, and the odds ratio (OR) or the relative risk of exposure $\mathbf{D(t)}$ can be of interest. In the current work, we focus on ordinal outcomes that are made of more than two categories, and use the marginal OR (or log-OR) as our \textit{population-average} quantity of interest. We next review the notation and statistical methodology that we will employ  for a general setting. 
 
\subsection{Assumptions and notation} \label{assum}

%To support that extension, we will rely on extended theory and simulation studies; Let put aside the application to the \textit{Add Health} study for a moment, and review the notation and the statistical methodology that we will use for estimation, for a more general setting.
Vectors and matrices are denoted in bold. Let $Y_i(t)$ be a longitudinal outcome (that is either observed or not) at time $t$, in individual $i$ ($i=1,...,n$), and assume it is categorical and it takes values in $\left\{ 1, 2, ..., J \right\}$. Let the exposure be denoted by $D_i(t)$ at time $t$, which is continuous and takes values in ${\rm I\!R^{+}}$. We are interested in the estimation of the marginal effect of an increment in exposure $D_i(t)$ on the outcome $Y_i(t)$. We suppose that the outcome is ordinal -- that is, it can be ordered such that categories $1$ and $J$ respectively represent the lowest, and the highest levels of that variable. \textcolor{black}{Further, the subsequent categories of the outcome are not considered to be equidistant, and going from e.g. category $j$ to $j+1$ may imply a more important change than going from category $j-1$ to $j$, etc.}

Further suppose that the outcome $Y_i(\cdot)$ of individual $i$ is only observed sporadically, at times denoted by $T_{i1},...,T_{iQ_i}$ (called \textit{monitoring} or \textit{visit} times). The quantity $Q_i$ denotes the number of observation times of individual $i$ in the time lapse $\left[0, \tau \right]$, with $\tau$ the maximum follow-up time in the whole study. As the indices indicate, the monitoring times and number of visits are allowed to vary across individuals. The monitoring indicator $dN_i(t)$ indicates whether there is ($dN_i(t)=1$) or is not ($dN_i(t)=0$) a visit of individual $i$ at time $t$. That indicator is expected to depend on certain patients' characteristics that we denote by the set $\mathbf{Z_i(t)}$. The set may contain the exposure of interest, $D_i(t)$, as well as mediators of the relationship of interest between $D_i(t)$ and $Y_i(t)$ or confounding factors. We denote the mediators of the relationship between $D_i(t)$ and $Y_i(t)$ by $\mathbf{M_i(t)}$; they are on the causal path from $D_i(t)$ do $Y_i(t)$. Confounders are denoted by $\mathbf{K_i(t)}$ and are assumed to affect both the outcome $Y_i(t)$ and the exposure $D_i(t)$ at time $t$ such that merely assessing the effect of the exposure on the outcome will lead to a distorted estimate.  
\begin{figure}[H]
\begin{minipage}{.45\textwidth}
\begin{center}
\begin{tikzpicture}[scale=0.7] 

\node (1) at (0,0) {$D_i(t)\subset \mathbf{Z_i(t)}$};
\node (2) at (6,-3) {$\mathbf{M_i(t)} \subset \mathbf{Z_i(t)}$};

 \node(3) at (6,-8){ $Y_i(t)$};
 \node(4) at (0.5,-7){$dN_i(t)$};
 \node(5) at (-2, -3){$\mathbf{K_i(t)}$};
 \draw[->](1) to (3);
 \draw[->](1) to (2);
  \draw[->](2) to (3);
\draw[->](5) to (1);
\draw[->](5) to (3);
\draw[->] (1) to (4);
\draw[->](2) to (4);
%\draw[->, dashed](5) to (4);
\end{tikzpicture} 
\end{center}
\end{minipage} \hfill
\begin{minipage}{.45\textwidth}
\begin{center}
\begin{tikzpicture}[scale=0.7] 

\node (1) at (0,0) {Video Games};
\node (2) at (6,-3) {Depressive mood};

 \node(3) at (6,-8){ Suicide attempts};
 \node(4) at (0.5,-7){Indicator Monitored};
 \node(5) at (-2, -3){Confounders \footnote{See Section \ref{appli} for the list of confounders.}};
 \draw[->](1) to (3);
 \draw[->](1) to (2);
  \draw[->](2) to (3);
\draw[->](5) to (1);
\draw[->](5) to (3);
\draw[->] (1) to (4);
\draw[->](2) to (4);
\draw[->](5) to (4);
\end{tikzpicture} 
\end{center}
\end{minipage}
\caption{The assumed causal diagram in the simulation studies (left panel) and the corresponding causal diagram for the application on the effect of video games on suicide attempts (right panel).} \label{fig1aa}
\end{figure}
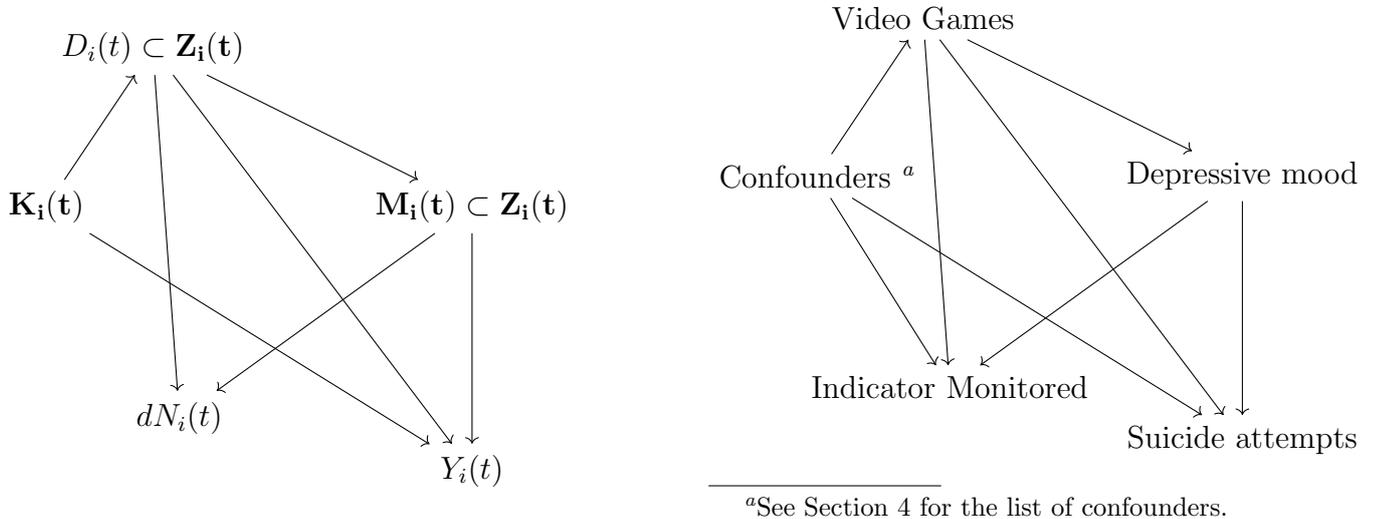

Figure \ref{fig1aa} (left panel) shows the causal diagram that we assume for the data generating mechanism discussed here, at each time $t$. Effectively, that diagram represents one \textit{slice} or \textit{snapshot} in time, but the relations depicted in that Figure are assumed to be true for each point in time as well as for each individual. This means that some covariates ($\mathbf{Z_i(t)}$ here) affect the monitoring propensity at each point in time, such that a prediction for $dN_i(t)$ would require the information on covariates $\mathbf{Z_i(t)}$ even on times $t$ $\subset \left[0, \tau \right]$ when $dN_i(t)=0$ and there is no visit. In that same causal diagram, the set $\mathbf{Z_i(t)}$ contains all covariates associated with $D_i(t)$ which themselves affect $dN_i(t)$; these are all the covariates affecting monitoring. In the right panel of Figure \ref{fig1aa}, we show the causal diagram that is posited to describe the data collected in the \textit{Add Health} study. We recall that, in the application, we are interested in the total causal effect of video games on the categorical suicide attempt outcome. This effect comprises the direct effect of exposure, but also that mediated through depressive mood.

Our aim is to build a pseudopopulation \citep{robins2000marginal} in which patients under different exposures are comparable with respect to confounding factors, and in which we adjust for any biasing path between the exposure $D_i(t)$ and the outcome $Y_i(t)$ that would be caused by conditioning on observed data (``visits''). To connect the causal and the statistical frameworks, we first need to reformulate the problem as a causal inference problem by using standard identifiability assumptions. To express the causal contrast of interest, we use the Neyman-Rubin potential outcome framework \citep{neyman1923application,rubin1974estimating} and denote by $Y_{id}(t)$ the potential outcome of individual $i$, at time $t$, and under exposure $d$. Necessarily, individual $i$ will only receive one level of the continuous exposure $D$ at time $t$, and therefore, only one of an infinite quantity of possible outcomes will be observed. The estimand we seek is the causal marginal OR for 1-unit increase in the continuous exposure $D_i(t)$. It is given by
\begin{align}
OR=  \left(\frac{\mathbb{P}\left[ Y_{id}(t) \le j   \right]}{1-\mathbb{P}\left[ Y_{id}(t) \le j  \right]}\right) \bigg/    \left( \frac{\mathbb{P}\left[ Y_{i(d+1)}(t) \le j   \right] }{1-\mathbb{P}\left[ Y_{i(d+1)}(t) \le j  \right]}\right)   \label{oddsratio}
\end{align}
and is constant across $j$. Note that the numerator in (\ref{oddsratio}) shows the potential outcome under exposure $d$, and the denominator, under exposure $d+1$, as we are presenting in (\ref{oddsratio}) the OR as a function of the probability for the outcome to be smaller or equal to a certain category $j$, rather than larger. Depending on the parameterization of the POM model, the estimand could vary.
  
The OR we aim to estimate requires assumptions regarding the data generating mechanism and its associated causal diagram. The causal diagram that we assume (presented in the left panel of Figure \ref{fig1aa}) is subject to both biasing paths due to backdoor paths (confounders), and to selection due to conditioning on monitoring indicators. These \textit{two} biases are due to the lack of exchangeability of the individuals across all exposure levels (\cite{bookcausalinference}, Chapters 7-8). For identifiability of the causal exposure effect, we must assume conditional exchangeability, consistency, as well as positivity of the exposure and the monitoring indicators, which are denoted by:
 \begin{center}
\begin{align}
\tag{P1} D_i(t) \perp  Y_{id}(t)   | dN_i(t), \mathbf{K_i(t)},\mathbf{Z_i(t)}  \hspace{0.3cm} \forall d  \in \mathbb{R^{+}} \label{p1} \\
\tag{P2} \text{If} \hspace{0.15cm} D_i(t)=d \hspace{0.1cm} \text{then}  \hspace{0.15cm}Y_{id}(t) = Y_i(t)  \\
\tag{P3} 0 < P(D_i(t)=d | \mathbf{K_i(t)})<1\hspace{0.2cm} \forall d \in \mathbb{R^{+}}\\
\tag{M1} 0 < P(dN_i(t)=1 | \mathbf{Z_i(t)})<1\hspace{0.2cm} \forall t \in [0,\tau].
\end{align}
\end{center}
 The assumption (\ref{p1}) means that the two sets (confounders $\mathbf{K_i(t)}$, and those affecting the monitoring times $\mathbf{Z_i(t)}$) are sufficient to block any biasing path from the exposure to the outcome of interest \textit{even} after conditioning on monitoring indicator $dN_i(t)$. \cite{bookcausalinference} (Chapter 8) presented a clear graphical representation via a probability tree, in the cross-sectional case, of the two types of biases which \textit{should} be considered simultaneously. Positivity for monitoring means that there is no time point (or time period, if time is considered to be continuous) when the probability that a visit will occur is null, or when it is 1 such that the visit is sure to occur. If such time point existed, this could lead to computational issues with the inverse weighting methodology that will follow, as well as concerns regarding the interpretation of findings.  In addition to the assumptions above, we assume that censoring times (or the individual times when patients' follow-up stops) are uninformative, in the sense that we capture through assumption \ref{p1} the only possible differences in follow-up that could ultimately bias our estimator for the effect of exposure on the outcome.

 \subsection{Methodology}

Following the assumptions we made in Section \ref{assum}, we can estimate the causal marginal OR using the POM. We now explain how we model the monitoring rate, the exposure, and how these models are combined to estimate the estimand.

For the monitoring model, we use an inverse intensity of visit (IIV) weight \citep{lin2004analysis} and model the intensity by using a proportional intensity model as a function of the covariates $\mathbf{Z_i(t)}$, which covariates (if considered together with the confounders) are assumed to create conditional exchangeability of the potential outcomes. The model is as follows:
\begin{align}
\mathbb{E}\left[ dN_i(t) | \mathbf{Z_i(t)}\right] = \xi_i(t) \exp(\bm{\gamma}'\mathbf{Z_i(t)}) \lambda_0(t) dt,\label{rate}
\end{align}
where $\xi_i(t)$ is an indicator for still being at risk, for individual $i$ at time $t$, and $\lambda_0(t) dt = d\Delta_0(t)$ with $\Delta_0(t)$ any non-decreasing function \citep{lawless1995some}. We use the Andersen and Gill model \citep{andersen1982cox}, an extension of the Cox proportional hazards model \citep{cox1972regression} to recurrent events, to estimate the parameters $\bm{\gamma}$ in (\ref{rate}). In this case, the ``recurrent events'' consist of the monitoring times. The baseline rate $\lambda_0(t)$ need not be estimated if it does not vary across individuals and if it is a function of the time since cohort entry (or time into study), rather than e.g.~time since the last visit. That is, because any function of time since cohort entry would be the same at time $t$ into the study across individuals, the function cancels out across individuals \citep{buuvzkova2009semiparametric}.

After adjusting for the monitoring intensity, under the mentioned assumptions, we can attain conditional exchangeability by fitting a correctly specified exposure model that is a function of covariates $\mathbf{K_i(t)}$, that will be used to break the links between the exposure and those covariates. However, we are interested in the case where exposure is continuous such that the standard inverse probability of treatment (IPT) weight cannot be implemented. To account for confounding under a continuous exposure, methods such as parametric g-computation were proposed \citep{ahern2009estimating}, however that method requires a correctly specified parametric outcome model, and it combines less naturally with our IIV weight. If the exposure distribution is (approximately) normally distributed, we propose to use a generalization of the standard IPT weight where we fit two linear models as proposed in \cite{robins2000marginal}: one where potential confounders are included as predictors, and another stabilizing model which only includes a constant intercept, as follows:
 \begin{align}
 \tag{L1}   \mathbb{E}[ D_i(t) | \mathbf{K_i(t)} ] =& \psi_0 + \bm{\psi_1}' \mathbf{K_i(t)} \label{pok0} \\
 \tag{L2}  \mathbb{E}[ D_i(t)] = & \psi_m.\label{pok}
 \end{align}
Once these models are fitted to the data, we obtain estimates for the parameters $\left\{ \hat{\psi}_0, \hat{\bm{\psi}}_1, \hat{\psi}_m \right\}$ and may compute the residuals from both models. %Residuals are computed as the difference in the fitted values for $D_i(t)$, and the actual exposure $D_i(t)$, for both models (L1) and (L2) and are respectively denoted by $\epsilon_{1,i}(t)$ and $\epsilon_{2,i}(t)$ for individual $i$ at time $t$.
 The generalized IPT weight, to be added to our weighted estimating equations, is then computed as
\begin{align}
\mathbf{e_i}(\mathbf{t}; \bm{\psi}) = \frac{h^{-1} (\hat{\psi}_0 + \hat{\bm{\psi}}_1' \mathbf{K_i(t)})}{h^{-1}(\hat{\psi}_m)}\label{ei}
\end{align}
for $h^{-1}(\widehat{D}_{l,i}(t)) = 1/\sqrt{2 \pi \widehat{\sigma}_l^2} \exp \left(- \widehat{\epsilon}_{l,i}(t)^2/(2 \widehat{\sigma}_l^2) \right)$ the Normal density function evaluated at the corresponding linear regression residuals $\widehat{\epsilon}_{l,i}(t)=\left(D_i(t)- \widehat{D}_{l,i}(t)\right)$, with $\widehat{\sigma}_l^2$ the  empirical variance of $\widehat{\epsilon}_{l,i}(t)$, $\widehat{D}_{l,i}(t)$ the model predictions, and $l=1,2$ the index corresponding to models (\ref{pok0}) and (\ref{pok}) respectively \citep{robins2000marginal}.  If the exposure is not approximately normally distributed (e.g., if its distribution is highly skewed), a log transformation of the exposure may lead to a more normally distributed (transformed) exposure. Other alternatives to the generalized IPT weight above include that where the exposure is binned in quantiles (e.g., deciles) and where a categorical model is fitted to estimate the probability for the exposure to belong to a respective category (as a function of covariates). \cite{naimi2014constructing} compared that option to the other weight mentioned above and they have found that the quantile binning approach performs better when the exposure is not normally distributed. \cite{schulz2020doubly} also discussed inverse weights in the context of a continuous dose for an exposure, when developing optimal adaptive dosing strategies. They considered binning the exposure in quantiles to compute the IPT weight, while leaving the exposure in its continuous form in the outcome model. They also used a similar quantiles approach as in \cite{naimi2014constructing} where they estimated the probability for the exposure to belong to a respective category using the POM model (rather than a conditional logistic regression model as used in \cite{naimi2014constructing}). They found that the binning approach can reduce the volatility in the weights. In simulation studies below, we demonstrate the proposed methodology with the generalized inverse weight in  (\ref{ei}) and in the case where the exposure is normally distributed. In the application to the \textit{Add Health} study, we assess the marginal effect of a 1-unit increase in the exposure (the number of hours spent playing video games) in its current form, and after it was log transformed. We assess the sensitivity of the results after using different inverse weighting strategies.

In this work, we focus on the estimation of the marginal OR for a 1-unit increase in the exposure $D_i(t)$ (or the log transformed exposure, in the application), with the odds being those of the outcome $Y_i(t)$. The OR was presented in equation (\ref{oddsratio}). Under the identifiability assumptions in Section \ref{assum}, we use a parametric POM model which models the mean outcome as follows:
  \begin{align}
P(Y_i(t) \le j | D_i(t) ) %= \frac{\exp(\alpha_j - \beta_D D_i(t))}{1+\exp(\alpha_j -  \beta_D D_i(t))} \nonumber \\
= \text{expit}(\alpha_j - \beta_D D_i(t)) \label{eqoutc}
\end{align}
$\forall j=1,...,J.$ Trivial algebra leads to the marginal OR:
\begin{align}
\frac{\left( \frac{\mathbb{P}\left[ Y_i(t) \le j | D_i(t) +1  \right] }{1-\mathbb{P}\left[ Y_i(t) \le j | D_i(t)+1  \right]}\right) }{ \left(\frac{\mathbb{P}\left[ Y_i(t) \le j | D_i(t)  \right]}{1-\mathbb{P}\left[ Y_i(t) \le j | D_i(t)  \right]}\right) } =& \frac{\frac{\exp(\alpha_j - \beta_D (D_i(t)+1) )}{1+\exp(\alpha_j - \beta_D (D_i(t)+1)} }{\frac{1}{1+\exp(\alpha_j - \beta_D (D_i(t)+1)} } \times  \frac{\frac{1}{1+\exp(\alpha_j -\beta_D D_i(t))}}{\frac{\exp(\alpha_j - \beta_D D_i(t))}{1+\exp(\alpha_j - \beta_D D_i(t))}}\nonumber \\  
=& \frac{\exp(\alpha_j - \beta_D (D_i(t)+1))}{\exp(\alpha_j - \beta_D D_i(t))}\nonumber \\
=& \exp(-\beta_D).\label{por}
\end{align}

\noindent Therefore, for making our causal inference, we are left with the estimation of the parameter $\beta_D$ in (\ref{por}). Using our models for exposure and monitoring, and re-weighting estimating equations by the IPT and the IIV weights, we create a pseudopopulation in which we have conditional exchangeability, so that the parameter $\beta_D$ can be estimated using directly the POM on the re-weighted data. That is, we extend the \textit{Flexible Inverse Probability of Treatment and Monitoring} weighted estimator \citep{coulombe} ($\hat{\beta}_{FIPTM}$) to the case where the exposure is continuous, and where the outcome mean is not assumed to be a linear function of covariates. The new, proposed, doubly-weighted estimator is further referred to as $\hat{\beta}_{IPTMP}$ for the \textit{Inverse Probability of Treatment and Monitoring POM} model.  

Denote by $\zeta_{i,j}(t)= P(Y_i(t) \le j  | D_i(t) )$ and by $Z_{i,j}(t)=\mathbb{I}(Y_i(t) \le j)$ with $\mathbb{I}(\cdot)$ the indicator function, which also correspond to the vectors
 \begin{align}
 \bm{\zeta}_i\mathbf{(t)}  = \left[ \begin{matrix} P(Y_i(t) \le 1  | D_i(t) ) \\  P(Y_i(t) \le 2  | D_i(t) ) \\ ... \\P(Y_i(t) \le J  | D_i(t) )  \end{matrix} \right] , \hspace{0.1cm}\text{and} \hspace{0.1cm}   \mathbf{Z_i(t)}= \left[ \begin{matrix} \mathbb{I}(Y_i(t) \le 1) \\ \mathbb{I}(Y_i(t) \le 2)  \\ ... \\ \mathbb{I}(Y_i(t) \le J)\end{matrix}\right],  
 \end{align}
where $\mathbf{Z_i(t)}$ is our modified outcome which accounts for the fact that the study outcome $Y_i(t)$ is made of several categories. Then, effectively, our methodology is equivalent to using the following estimating equation (an exension of \cite{lin2004analysis} and \cite{coulombe}) to estimate the marginal effect of exposure via the log-OR for 1-unit increase in exposure, denoted by $\beta_D$:

\begin{align}
\mathbb{E} \left[ \int_0^{\tau}\frac{
 \mathbf{e}(\mathbf{t}; \bm{\psi}) \left(  \mathbf{Z(t)} -  \bm{\zeta}\mathbf{(t)} \right)  } {   \bm{\varphi}\mathbf{(t; \bm{\gamma})   }}   \mathbf{dN(t)} \right]=\mathbf{0}, \label{eq}
\end{align}

\noindent where $\mathbf{e}(t; \bm{\psi}) $ is our generalized IPT weight for a continuous exposure, $ \bm{\varphi}\mathbf{(t; \bm{\gamma})}$ an IIV weight for the monitoring, and where the parameterization of $ \bm{\zeta}\mathbf{(t)}$ is 
\begin{align}
\zeta_{i,j}(t) = & \text{expit}(\alpha_j - \beta_D D_i(t)).
 \end{align}
  
\noindent For the inverse intensity of visit function, we simply plug in our estimated model from (\ref{rate}); that is
\begin{align*}
\bm{\varphi_i}\mathbf{(t; \widehat{\bm{\gamma}})} = \exp \left(  \widehat{\bm{\gamma}}'\mathbf{Z_i(t)}   \right),
\end{align*}
where the parameters $\bm{\gamma}$ can be estimated using the \textit{coxph} function in R, from the \textit{survival} package \citep{survival-package}. The baseline rate in (\ref{rate}) need not to be estimated, as underlined earlier. For the IPT weight, we use an estimate of (\ref{ei}) that is computed by fitting both linear models and computing the respective residuals as discussed earlier (and assess a few other weighting options in the application to the \textit{Add Health} study). The weighted POM can be fitted using, for instance, the \textit{polr} function from the \textit{MASS} package in R \citep{MASSpackage}.

\section{Simulation study}
We conducted several simulation studies to assess the proposed methodology in a setting where, in contrast to the \textit{Add Health} study, monitoring times can occur at any time during follow-up, for every individual. Our aim was to estimate the causal marginal OR for a 1-unit increase in the exposure $D_i(t)$ on a categorical and ordinal outcome $Y_i(t)$.  The outcome was categorical, taking one of three levels ($J=3$): 1, 2, and 3.  In simulation studies, we compared four estimators:
\begin{itemize}
\item The estimated log-OR for exposure obtained directly from the POM model, with no adjustment ($\hat{\beta}_{POM}$);
\item The estimated log-OR for exposure from a weighted POM with an IPT weight, where the propensity score is a correctly specified function of the confounders ($\hat{\beta}_{IPTP}$);
\item The estimated log-OR for exposure from a weighted POM with an IIV weight, where the intensity is a correctly specified function of the covariates affecting visit times ($\hat{\beta}_{IIVP}$);  and
\item The estimated log-OR for exposure from a doubly-weighted POM with both the IPT and the IIV weights, with both functions (corresponding to the exposure and visit models) correctly specified  ($\hat{\beta}_{IPTMP}$).
\end{itemize}

In the following description of the data generating mechanism that we used, the individual index is omitted for ease of exposition. We used 1000 simulations per study and tested settings with either 250 or 1000 patients per simulated dataset. First, three confounders were simulated at time 0 (``cohort entry'') and for each individual, as $K_1 \sim N(1,1)$, $K_2 \sim Bernoulli(0.55)$, and $K_3 \sim N(0, 1)$. The time-varying, continuous exposure $D(t)$ was simulated at each time $t$ from a Normal distribution with a mean that depended on the confounders, as $D(t) \sim N(-0.5 + 0.5 K_1 + 1 K_2 -0.05 K_3; 0.5^2)$ in the study with confounding, and $D(t) \sim N(-0.5; 0.5^2)$ in the simulation study with no confounding.  The mediator of the relationship between $D(t)$ and $Y(t)$ was binary and time-varying, and simulated as $Z(t)\sim Bernoulli(p_D(t))$ with $p_D(t) = 0.3$ if $D(t) >0.5$ and $p_D(t) = 0.8$ otherwise.

In the study, time was continuous, and discretized over a grid of 0.01. Visit times could vary across individuals, and could occur at anytime during follow-up. %, as opposed to our application to the \textit{Add Health} study. 
Monitoring times were simulated according to the monitoring intensity, %simulated 
as a function of $D(t)$ and $Z(t)$ and with an individual random effect $\eta$, such that $\lambda(t|  D(t), Z(t)) = 0.01 \eta \exp\left( \gamma_D D(t) + \gamma_Z Z(t) \right)$. We varied the parameters $\bm{\gamma}$ to obtain different strengths for the dependence between the covariates and the monitoring times. The random effect was simulated as a random Gamma variable with mean 1 and variance 0.01. Monitoring indicators at each time were simulated according to a random Bernoulli draw, with a probability proportional to the intensity above. 

    The categorical outcome $Y(t)$ was simulated as a function of the exposure $D(t)$, the mediator $Z(t)$, and the confounders $\mathbf{K(t)}$. We first simulated the linear mean function as $\mu(t| D(t), Z(t), \mathbf{K(t)}) = -2 D(t) + 5 Z(t) + 0.4 K_1 + 0.05 K_2  -0.6 K_3$. Then, a random draw of the logit probability was simulated, according to the mean $\mu(t| D(t), Z(t), \mathbf{K(t)})$. If that value was smaller or equal to 5, the categorical outcome was set to 1. If the value was greater than 5 and smaller or equal to 8, it was set to 2. It was set to 3, otherwise. That choice of thresholds led to a good distribution across all three levels.

 Given that we used the POM to estimate the marginal effect of exposure (which assumes that the relation between the linear predictors and the outcome, the link function, is the expit function), and given that some mediators make up part of the total effect of exposure on the outcome, the true marginal log-OR cannot be analytically derived solely by knowing the simulation parameter for the exposure in the outcome model. To know the true value (or \textit{target} for an estimator) we conducted a Monte Carlo simulation in which we simulated the data of 10,000 patients a total of 1000 times where all parameters in the outcome model were kept as above, but there was no covariate-driven treatment or visit process such that there was no selection or confounding bias. We then computed the log-OR of exposure each time. In that very large study, no imbalances due to confounding or due to covariate-driven monitoring times were present. For that, we merely set all the parameters corresponding to the predictors in the exposure and in the monitoring models to zero. We obtained the target marginal log-OR by averaging the 1000 log-OR for exposure across those simulations, which equalled to -1.061.

 \subsection{Results}
 
\begin{figure}[H]
    \begin{minipage}{.43\textwidth}
    \centering
    \includegraphics[width=1.\textwidth,valign=t]{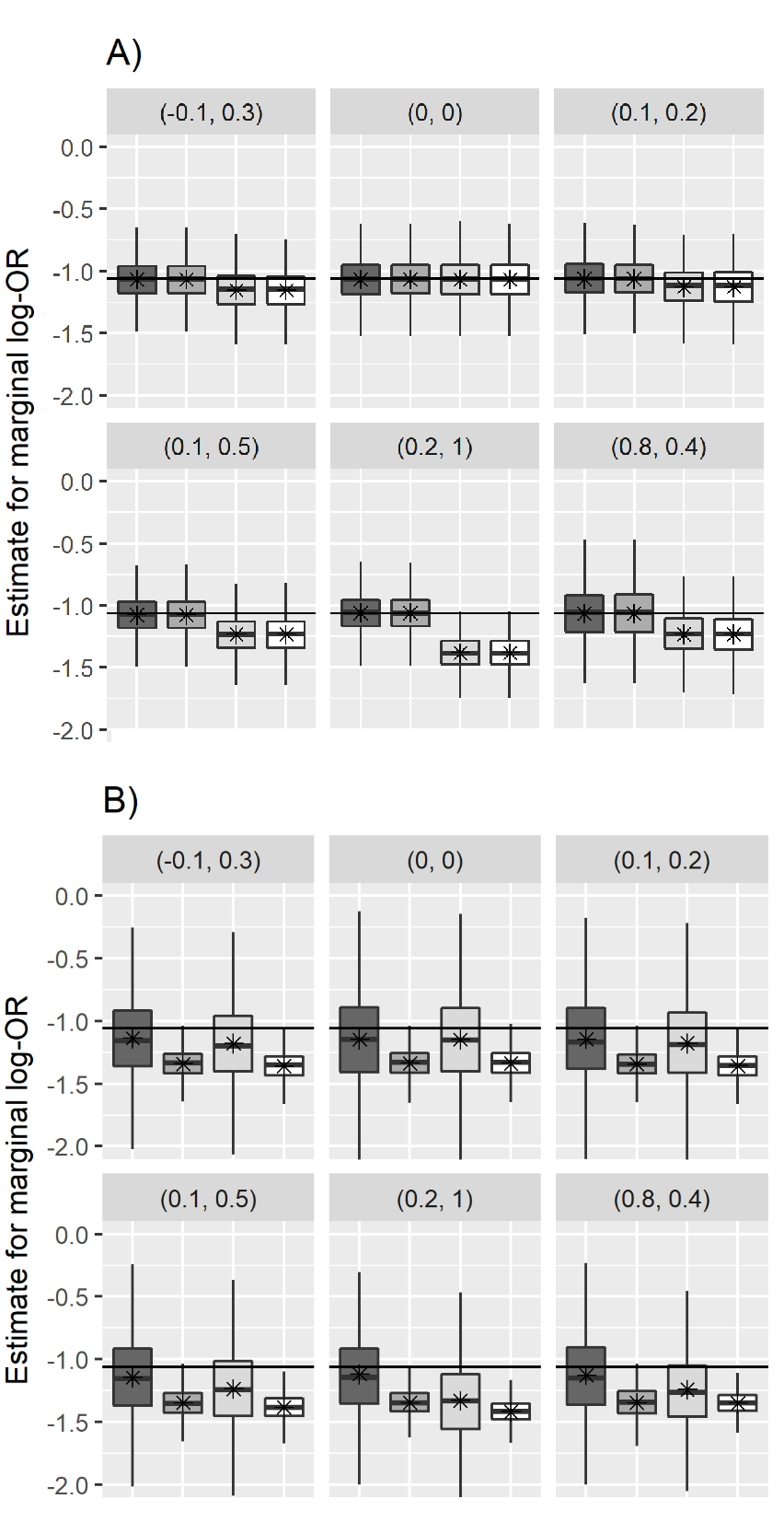}
    \end{minipage} \hspace{0.75cm}   
        \begin{minipage}{0.45\textwidth}
    \centering
  \includegraphics[width=1.25\textwidth,valign=t]{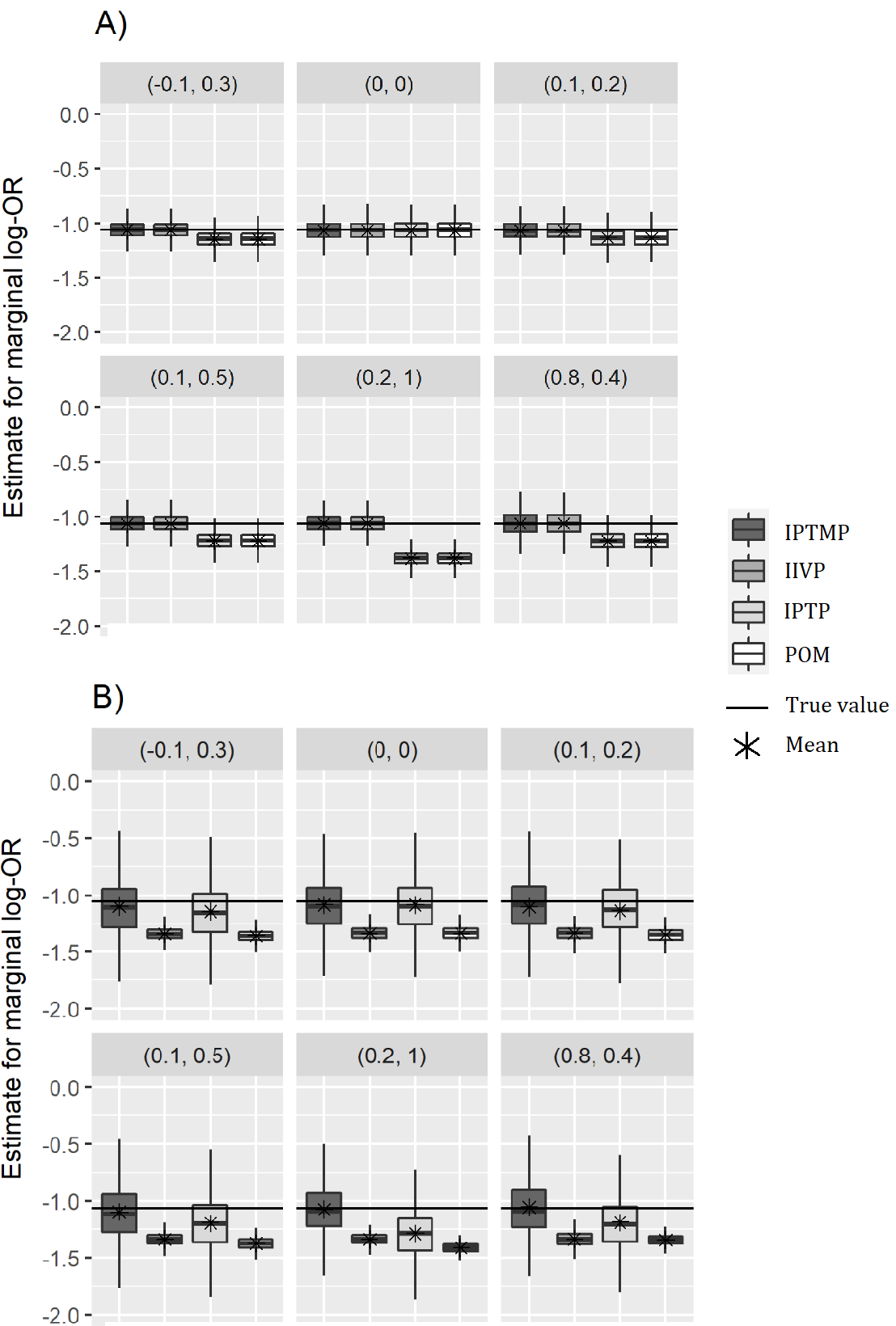}
   \end{minipage}
\par
\begin{minipage}[t]{.43\textwidth}
\centering
  \caption*{250 patients}  
\end{minipage}\hspace{0.75cm}    
\begin{minipage}[t]{.45\textwidth}
\centering
  \caption*{1000 patients}  
\end{minipage}  
  \caption{ A) Simulation study with no confounding. B) Simulation study with confounding. Study with 250 patients (left panel) or 1000 patients (right panel). Distribution of the estimated marginal log-OR for a 1-unit increase in the continuous exposure $D_i(t)$ (1000 simulations per study) across different scenarios for the monitoring process ($\bm{\gamma}$ parameters at the top of each subgraph) and for the 4 estimators compared. The horizontal dark line represents the true value of the effect. The left panel values correspond to those reported in Table \ref{tab:var}. }\label{tresults}
    \end{figure}

  Figure \ref{tresults} shows the results of the simulation study for two settings (250 patients in the left panel, and 1000 patients in the right panel), and for the case where there is no confounding (top panel) and where there is confounding (bottom panel). As expected, the proposed doubly-weighted estimator $\hat{\beta}_{IPTMP}$ was the least biased across all four estimators. In the case with no confounding, its performance was equivalent to that of $\hat{\beta}_{IIVP}$, which also accounted for covariate-driven monitoring times. In the case with confounding, the data presented with biasing imbalances related to both the confounders and the covariate-driven monitoring times, and $\hat{\beta}_{IPTMP}$ was the least biased across the board (it was the only estimator to account for both of these features). When increasing the sample size, our proposed estimator converged to the true value (horizontal dark line in Figure \ref{tresults}) while all other estimators remained biased.

 \begin{table}[H]
 \begin{center}
\caption{Comparison of four estimators for the marginal log-OR for 1-unit increase in $D_i(t)$ in the POM, for a sample size of $n=250$ patients and 1000 simulations per study. Study without confounding (Conf.=N) and with confounding (Conf.=Y).}
\begin{tabular}{cc c c c cc c c cc}
  \hline  \hline
Conf. &$\bm{\gamma}$ & Mean no.&  \multicolumn{4}{c}{Absolute empirical bias} &   \multicolumn{4}{c}{Empirical variance}\\
  (Y/N)&  & visits & IPTMP & IIVP  & IPTP & POM& IPTMP & IIVP  & IPTP & POM\\ 
  &&(min-max)&&&&&&&& \\
  \hline
 N &(-0.3, 0.1)  & 3 (0-14) &  0.01& 0.01 & 0.02 & 0.02 & 0.02 & 0.02& 0.02& 0.02                     \\ 
   &(-0.1, 0.3)  & 3 (0-15) &   0.01&0.01&0.09&0.09&0.03&0.03&0.03&0.03                    \\      
   &(0, 0)	   & 2 (0-13)    &   0.01&0.01&0.01&0.01&0.03&0.03&0.03&0.03                      \\      
   &(0.1, 0.2)   & 3 (0-14) &    0.00 & 0.00 & 0.07 & 0.07 &0.03&0.03&0.03&0.03             \\      
   &(0.1, 0.5)   & 3 (0-15) & 0.01&0.01&0.17&0.17&0.03&0.03&0.03&0.03                     \\      
   &(0.2, 1)      & 5 (0-19) & 0.00&0.00& 0.32&0.32 &0.02&0.02&0.02&0.02              \\      
   &(0.8, 0.4)   & 2 (0-12)  &0.00&0.00& 0.17&0.17 &0.05&0.05&0.03&0.03 \\      \hline
 Y &(-0.3, 0.1)  & 2 (0-13) &  0.09 & 0.28 & 0.11 & 0.27 & 0.24 & 0.02 & 0.24 & 0.02                     \\ 
   &(-0.1, 0.3)  & 3 (0-16) &   0.08& 0.28 & 0.12 & 0.30 & 0.19&0.01&0.19 & 0.01                     \\      
   &(0, 0)	   & 2 (0-13) &     0.09 & 0.28& 0.09 & 0.28 & 0.23 & 0.02 & 0.23 & 0.02             \\      
   &(0.1, 0.2)   & 3 (0-16) &   0.09 & 0.28 & 0.12 & 0.30 & 0.21 & 0.01 & 0.20 & 0.01               \\      
   &(0.1, 0.5)   & 3 (0-15) &     0.08 & 0.28 & 0.17 & 0.32 & 0.17&0.01&0.17&0.01                   \\      
   &(0.2, 1)      & 5 (0-21) &    0.06 & 0.29 &0.27 & 0.36 &0.16&0.01&0.14&0.01              \\      
   &(0.8, 0.4)   & 5 (0-46) &     0.07 & 0.28 & 0.18&0.29 &0.16 &0.02 &0.14 &0.01               \\          \hline
\end{tabular}
 \label{tab:var}
 \end{center}
 \end{table}

 %  \begin{figure}[H]
  %\includegraphics[width=17.5cm]{no_confounding2.png}
  %\end{figure}
 
\noindent   We show a comparison of the empirical variance of all four estimators in Table \ref{tab:var}. There, we find that the proposed estimator $\hat{\beta}_{IPTMP}$ is approximately as variable as the IPT-weighted estimator, with both being more variable than other estimators compared.   \textcolor{black}{However, given that the proposed estimator $\hat{\beta}_{IPTMP}$ is unbiased, the empirical mean squared error (MSE) for that estimator was smaller across the board, in settings with no confounding (results not shown). In settings with confounding, even if less biased, the MSE for $\hat{\beta}_{IPTMP}$ was typically greater than that of $\hat{\beta}_{IIVP}$, indicating that the variance due to the generalized IPT weight was quite important. As expected, the gap between the empirical MSE of $\hat{\beta}_{IPTMP}$ and $\hat{\beta}_{IIVP}$ in the confounded setting tended to decrease as the sample size grew (the MSE of $\hat{\beta}_{IPTMP}$ was 112\% larger on average, in the case with $n=250$, and 45\% larger when $n=1000$). In settings with more confounding, or if we increased the sample size in our studies, we therefore would expect smaller MSEs across the board with the proposed $\hat{\beta}_{IPTMP}$, while other estimators would remain biased.} %Stabilizing the weights could possibly help in reducing the variance of the proposed estimator (see e.g. \cite{robins2000marginal} for a discussion on stabilization for the exposure model, and \cite{buuvzkova2009semiparametric} for stabilization discussed in the context of inverse intensity of visit weight).

\section{Application to the \textit{Add Health} study} \label{appli}

The proposed methodology was applied to the \textit{Add Health} study, a longitudinal study made of several waves, in which individuals were followed from their adolescence to their adulthood \citep{harris1}. Individuals' personal data were collected via several questionnaires. They included demographics, social and biological determinants, behavioral determinants, and others. For the purpose of this analysis, we used public-use data sets from the first four waves, which are free and available online \citep{harris3}. These data respectively correspond to years 1994-1995, 1996, 2001-2002, and 2008. We solely focused on the data obtained from the in-home questionnaires, which were available at all four waves but that did not contain all the same questions, as well as the parent questionnaire (available for the first wave only). The sample population contained 6504 individuals.

Our aim was to assess the (causal) marginal effect of the number of hours spent gaming with video and/or computer games (further referred to as video games) per week, on the number of suicide attempts. In all four in-home questionnaires (corresponding to the four waves), the question \textit{During the past 12 months, how many times did you actually attempt suicide?} was asked, and responses were categorized differently across waves (Waves 1-2: 0 time, 1 time, 2-3 times, 4-5 times, 6 or more times; Waves 3-4: 0 time, 1 time, 2 times, 3-4 times, 5 or more). To have a consistent outcome definition across all waves, the study outcome was further categorized as 0 attempts, 1 attempt, or 2 or more attempts at suicide. In all four in-home questionnaires, individuals were also asked \textit{How many hours a week do you play video or computer games?}. This variable was used to define the exposure. Given that the variable distribution was highly skewed, in the main analysis, we log (base 2) transformed the exposure (after adding one unit), yielding a more symmetric and approximately normally distributed exposure. The log base 2 would also provide a straightforward interpretation for a 1-unit increase in the transformed variable, which then corresponded to a 2-fold increase in the former \textit{skewed} variable, the number of hours spent playing video games. Four individuals who respectively answered that they played video games 120, 140, 168 and 168 hours per week at one of the four waves had that number truncated to 112 hours per week before conducting the log transformation, a maximum that corresponded to an average of 16 hours of gaming per day. We postulated that the effect of video games on suicide attempts is mediated by the depressive mood (see Figure \ref{fig1aa}, right panel). Information on individuals' mood was available via the question \textit{How often was the following true during the past week? You felt depressed.} Possible answers to that question were consistent across all waves, and consisted in an ordinal scale from 0 to 3, with 0 being never or rarely, and 3 being most of the time or all the time.

For the causal question of interest, we assume that exposure to video games (possibly on a transformed scale) affects the marginal odds of suicide attempt both directly, and indirectly via the depressive mood \citep{goldfield2016screen,maras2015screen,johnson2013videogames}, which therefore lies on the causal path from the exposure to video games, to the outcome. Other mediators, which are affected by exposure to video games and could affect suicide, such as the quality of being fearless about death or having elevated physical pain tolerance \citep{houtsma2017longitudinal}, were not available in this particular study. Further, we postulate that being depressed and/or spending time on video games, as well as other characteristics in the confounder set, may make a participant less willing to complete a questionnaire (therefore influencing their chances of being ``monitored'' according to our definition). These assumptions are depicted in the causal diagram in Figure \ref{fig1aa}, right panel. 

The set of potential confounders included age, sex, socioeconomic status (SES) (defined using two questions asked to one of the participant's parent: \textit{About how much total income, before taxes did your family receive in 1994?} and \textit{How far did you go in school?}; the answers were transformed into quintiles and summed to give a score betwen 0 and 10, with 10 the highest SES), ethnicity, the frequency of having trouble relaxing (FHTR), the level of grooming of the respondent (LGR), seeing that the respondent seemed bored or impatient (RSBI), their most recent grades in mathematics (MATH), English or language arts (ENG), History or Social Sciences (HSS) and in Science (GS), the frequency with which they hang out with friends (HOF), their feeling that friends cared about them (FCA) and the number of days when they smoked cigarettes over the past month. The waves when these covariates were measured, and therefore could potentially be updated, are described in Table \ref{tab00} under the \textit{Exposure model} column. 

In the study, visits (or monitoring times) consisted of the times or waves when the outcome related to suicide attempts were available. We chose a set of predictors for the monitoring times which reflected our beliefs about which individuals' characteristics can influence both their response availability and the number of suicide attempts. Our choice was also influenced by the work of \cite{kalsbeek2002predictors} on determinants of nonresponse in the \textit{Add Health} study. For the monitoring model, we therefore selected age, sex, SES, ethnicity, the variables FHTR, LGR, RSBI, MATH, ENG, HSS, GS, HOF, FCA, cigarette consumption in the past month, as well as the weekly number of hours spent playing video games, and, as defined earlier, the frequency of feeling depressed, which is a potential mediator of the effect under study. The waves when these covariates were measured are also described in Table \ref{tab00}, under the \textit{Monitoring model} column. 

Since the predictors of monitoring must be measured both when there is ($dN_i(t)=1$) and there is no visit ($dN_i(t)=0$) to properly model the monitoring indicators, multiple imputation with 5 replicated datasets \citep{rubin1976inference} was used to impute missing covariates on times when there was no visit, as well as on times when the outcome was \textit{not} missing but that these variables were simply not recorded or assessed. Before conducting the imputation, some variables were merely replaced by sensible summary values: the grades (MATH, ENG, HSS, GS), which were only measured in the first two waves, were averaged over those two waves and used to replace missing values at any of the four waves. The sex, SES, and ethnicity, were defined only once at baseline and duplicated at all other waves. If they were missing in Wave 1, they were imputed separately at all four waves based on other characteristics. Age at all four waves could be inferred by using the years corresponding to each wave, and the year of birth for each individual. It was only missing if no information on date of birth or age was ever available. After imputation, all covariates (including the exposure and the mediator of interest) were completely filled, except for the outcome that was left as is. More details on the procedure we used for imputation, the rates of missing values in each covariates, and the performance of the imputation can be found in Appendix A.

\begin{table}[H]
 \begin{center}
\caption{Variable definition for the analysis of the \textit{Add Health} study, United States, 1994-2008, $n=6504$ individuals. Waves 1, 2, 3 and 4 are respectively represented by acronyms W1, W2, W3, and W4. A column presents the times (waves) when these questions were asked to participants or their parents. }
\begin{tabular}{ l ccc}
 \hline  \hline
Variable & Times of  & Exposure   & Monitoring   \\
  (\textit{question}) &  measurement &   model &   model \\
  \hline
Age & W1, W2, W3, W4 & X & X \\
Sex & W1$^\dagger$ & X &X \\ 
Socioeconomic status (SES)$^{\ddagger}$ &W1$^\dagger$ &X& X\\
Ethnicity&W1$^{\dagger}$&X&X\\
Frequency of having trouble relaxing (FHTR)$^\zeta$& W1, W2&X&X \\
Level of grooming of the respondent (LGR)$^\nu$ &W1, W2, W3, W4 &X&X \\
Respondent seemed bored or impatient (RSBI)$^\nu$ &W1, W2, W3, W4 &X&X \\
Most recent grade in Mathematics (MATH) &W1, W2&X&X \\
Most recent grade in English/language arts (ENG)&W1, W2&X&X \\
Most recent grade in History/Social sciences (HSS)&W1, W2&X&X \\
Most recent grade in Science (GS) &W1, W2&X&X \\
Frequency of hanging out with friends (HOF)$^\iota$&W1, W2, W3&X&X\\
Feeling that friends care about you (FCA)&W1, W2&X&X\\
How many days of smoking cigarettes over past month &W1, W2, W3, W4&X&X \\
Number of hours spent on video or computer games&W1, W2, W3, W4& &X\\
Frequency of feeling depressed &W1, W2, W3, W4& & X\\  \hline
\end{tabular} \label{tab00}

 \scriptsize{$\dagger$ Considered as remaining fixed throughout the study; $\ddagger$ Defined as the decile of a combination between patients' salary and patients' education; $\nu$ Question answered by the researcher questioning the participant, rather than directly by the respondent; $\iota$ In the past week} 
 \end{center}
  \end{table}
  Separate proportional intensity models for monitoring, and separate linear regression models for the exposure (one as a function of the confounders, and another for stabilization), were fitted on each imputed dataset. In those models, we used the covariates mentioned above and presented in Table \ref{tab00}. Time since cohort entry (thus, since Wave 1) was the time axis we considered in the Andersen and Gill model for the monitoring rates, and therefore the baseline monitoring rate $\lambda_0(t)$ canceled across individuals at each wave, such that it did not require estimation.

\begin{table}[H]
 \begin{center}
\caption{Estimated rate ratios (95\% CI) for the monitoring model, \textit{Add Health} study, United States, 1994-2008, $n=6504$ individuals. }
\begin{tabular}{ l c}
 \hline  \hline
Variable & Rate ratio (Bootstrap 95\% CI)\\ \hline
 Number of hours spent on video or computer games& 1.00 (1.00, 1.00)\\
Frequency of feeling depressed (Ref.= Never or rarely) &  \\
\hspace{0.4cm} Sometimes & 1.00 (0.97, 1.02)\\
\hspace{0.4cm} A lof of the time & 0.99 (0.94, 1.03)\\
\hspace{0.4cm} Most of the time or all the time &1.01 (0.94, 1.08) \\
Age& 0.93 (0.93, 0.94) \\
Sex (Female)& 1.11 (1.09, 1.13)\\
SES& 1.01 (1.01, 1.02)\\
Race (Ref.= White)& \\
\hspace{0.4cm} Black/African American & 0.93 (0.91, 0.95)\\
\hspace{0.4cm} American Indian/Alaskan Native & 0.96 (0.87, 1.04)\\
\hspace{0.4cm} Asian/Pacific Islander & 0.92 (0.88, 0.97) \\
\hspace{0.4cm} Other& 0.90 (0.86, 0.94)\\
FHTR &1.00 (0.99, 1.02) \\
LGR& 0.99 (0.98, 1.00)\\
RSBI& 0.98 (0.95, 1.02)\\
MATH& 1.00 (0.99, 1.01)\\
ENG& 1.00 (0.99, 1.01)\\
HSS& 1.00 (0.99, 1.01)\\
GS&  0.99 (0.98, 0.99)\\
HOF& 1.00 (0.99, 1.02)\\
FCA& 1.00 (0.98, 1.01)\\
How many days of smoking cigarettes over past month& 1.00 (1.00, 1.00) \\
    \hline
\end{tabular}
 \label{tab1}

 \end{center}
 \end{table}
 Average rate ratios and estimates for the marginal log-OR of exposure (the log base 2 transformed number of hours spent playing video games) were computed using Rubin's rule for multiply imputed datasets \citep{rubin2004multiple}. For computing 95\% confidence intervals (CI), we used a nonparametric bootstrap. Data were sampled with replacement within each participant cluster, so that each individual had the same number of monitored outcomes in each sample as in the original dataset and within-person correlation was maintained across bootstrap resamples.
 
We conducted three additional analyses to assess the sensitivity of the results to non-normality of the exposure distribution. First, in analysis S1, we assessed the original methodology with the generalized IPT weight presented in equation (\ref{ei}) to estimate the log-OR for a 1-unit or a 10-hour increases in the number of hours spent playing video games weekly (not transformed). Second (S2), the same exposure variable as in analysis S1 was used, but the IPT weight was computed by categorizing the exposure in five bins and using a POM model to fit the estimated probability of belonging to a respective category as a function of potential confounders. That probability was then used in an inverse weight in the weighted estimating equations. The five categories for the exposure were: 0 hours, 1 hour, and three other ranges based on the tertiles of the rest of the exposure distribution (the number of hours spent playing video games). Those categories were chosen for the high frequencies of 0 and 1 in the number of hours spent gaming in the dataset. Third (S3), the same analysis as in S2 was reproduced, but the exposure in the outcome model was the log base 2 transformed number of hours spent playing video games, rather than the number of hours itself. The categorical IPT weight was used in S3, with each observation being assigned the same category as in S2 (given there is a one-to-one mapping between the continuous and the log base 2 transformed variables). For S3, contrary to S1 and S2, we looked at the marginal effect of a 1-unit increase or 3-unit increase in the log base 2 number of hours, which respectively correspond to a 2-fold or 8-fold increases in the number of hours spent playing video games.

 The estimated rate ratios for monitoring are shown in Table \ref{tab1}, along with bootstrap 95\% CI. We found that characteristics such as being older, being male, lower SES, being Black/African American, Asian/Pacific Islander, or from another race than those listed, as well as having a greater grade in science, were statistically significantly associated with a lower chance of having replied to the question on suicide attempts. Recall that the group that did not respond to the question was a mix of patients who did not respond to any questionnaire at a given wave (who were completely absent from the study), and those who simply did not respond to the question on suicide attempts in particular. 
 
In the main analysis, the four estimators we compared for the marginal OR for a 1-unit increase or 3-unit increase in the log base 2 number of hours spent on video games are presented in Table \ref{tab2}, with their respective bootstrap CI. A 1-unit or a 3-unit increases in the log base 2 number of hours respectively correspond to a 2-fold or a 8-fold increases in the number of hours spent playing video games. We find in Table \ref{tab2} that most estimators show a decrease in the number of suicide attempts for a greater number of hours spent on video games (2-fold OR: 0.91 ($\hat{\beta}_{POM}$), 0.95 ($\hat{\beta}_{IIVP}$)) while our doubly-weighted estimator $\hat{\beta}_{IPTMP}$ provides an estimate for a 2-fold increase in the number of hours of gaming that corresponds to a multiplicative effect of 1.05 (95\% CI 0.92, 1.15) on the odds of passing to the next category of the categorical suicide attempts variable (i.e., going from 0 to 1 suicide attempt or from 1 to 2 or more attempts).

\begin{table}[H]
 \begin{center}
\caption{Comparison of four estimators for the marginal OR for a two-fold or 8-fold increases in the time spent on video games per week, on the odds of suicide attempts   (number of attempts categorized in 0, 1, or more), \textit{Add Health} study, United States, 1994-2008, $n=6504$ individuals. Confidence intervals computed via bootstrap resampling.}
\begin{tabular}{ l c  c}
 \hline  \hline
Estimator & 2-fold increase OR (95\% CI)  & 8-fold increase OR (95\% CI)  \\ \hline
 $\hat{\beta}_{POM}$ &0.91 (0.83, 0.98)& 0.76 (0.57, 0.96)\\
$\hat{\beta}_{IPTP}$ & 0.99 (0.89, 1.08)& 0.98 (0.69, 1.28) \\
$\hat{\beta}_{IIVP}$ &0.95 (0.85, 1.03) & 0.86 (0.61, 1.09)\\
$\hat{\beta}_{IPTMP}$ &1.05 (0.92, 1.15)& 1.15 (0.78, 1.53)  \\
    \hline
\end{tabular}
 \label{tab2}
 \end{center}
 \end{table}
 
The coefficients (log-OR) were not statistically significantly different from 0 for the effect of video games. Our proposed doubly-weighted estimator $\hat{\beta}_{IPTMP}$ was the only estimator showing a (non-significant) increase in the probability of more suicide attempts when playing more video games. The three sensitivity analyses led to similar results (Supplementary Material A, Supplementary Tables 5 to 7). Using the coefficients for the marginal OR of exposure and the outcome category-specific intercepts estimated with all four estimators, we can estimate the probability of 1 or more suicide attempt(s) for any value of the log base 2 transformed number of hours spent playing video games, or for the corresponding number of hours spent playing video games. For instance, in the main analysis, we estimated this probability for 5, 10, 30, 70, or 100 hours spent playing video games, and respectively obtained probabilities of (2.3, 2.1, 1.9, 1.7, 1.6) with the POM estimator, (2.0, 1.9, 1.8, 1.7, 1.6) with the IIVP estimator, (2.6, 2.6, 2.6, 2.6, 2.5) for the IPTP estimator, and (2.3, 2.4, 2.6, 2.7, 2.8) with the IPTMP estimator. In Figure \ref{figgh}, we plot and compare the estimated marginal probability of 1 or more suicide attempt given the time spent on video games, as well as the marginal probability of 2 or more attempts. The Figure also includes a rug plot on the X-axis, that shows the values that the exposure (the number of hours spent on video games) takes in the dataset, up to 110 hours per week. Four observations with respectively 120, 140, 168 and 168 hours of gaming per week are not included in that rug plot. In Supplementary Material B, we present these estimated probabilities along with the corresponding 95\% CI computed from the percentiles of the bootstrap distribution (Supplementary Figure 4). The bootstrap sampling procedure accounted for the variance of all respective fitted coefficients used to compute the probabilities. We also present the same plots that correspond to each of the three sensitivity analyses, either with or without the corresponding 95\% CI -- which tend to crowd the graphs -- (Supplementary Figures 5 to 10). We observed similar trends as in the main analysis in all these plots. 

While it has been shown that the exposure to active or to serious video games could be beneficial to teenagers (see e.g.~\cite{zayeni2020therapeutic}, \cite{zurita2018effect}), most studies on that topic were cross-sectional studies that did not look at longitudinal effects of video games. Further, it is yet not clear that non-active video games are not detrimental to individuals' health and wellbeing (see e.g.~\cite{teismann2014influence}, \cite{messias2011sadness}, and \cite{anderson1986affect}). We hereby have found a non-significant detrimental effect of increasing the amount of time spent on video games each week when accounting for bias due to confounding and selection to report. That effect is estimated marginally but there could be differential effects across sex (as suggested in \cite{anderson1986affect}) or by the type of video games. That latter feature, in addition to some other social behavior determinants or parental determinants, were not available to us, and therefore the marginal effect observed is one that may combine the effects in several different subgroups of video games users (active, non-active, addicted, etc.) which differ on characteristics. Finally, we could not find similar studies that were experimental and in which exposure to video games was randomized. More research is needed, with larger datasets, to understand whether there can be a significant detrimental effect of a large amount of time spent on video games weekly, on suicide attempts. 
  \begin{figure}[H]
   \begin{center}
  \includegraphics[width=13cm]{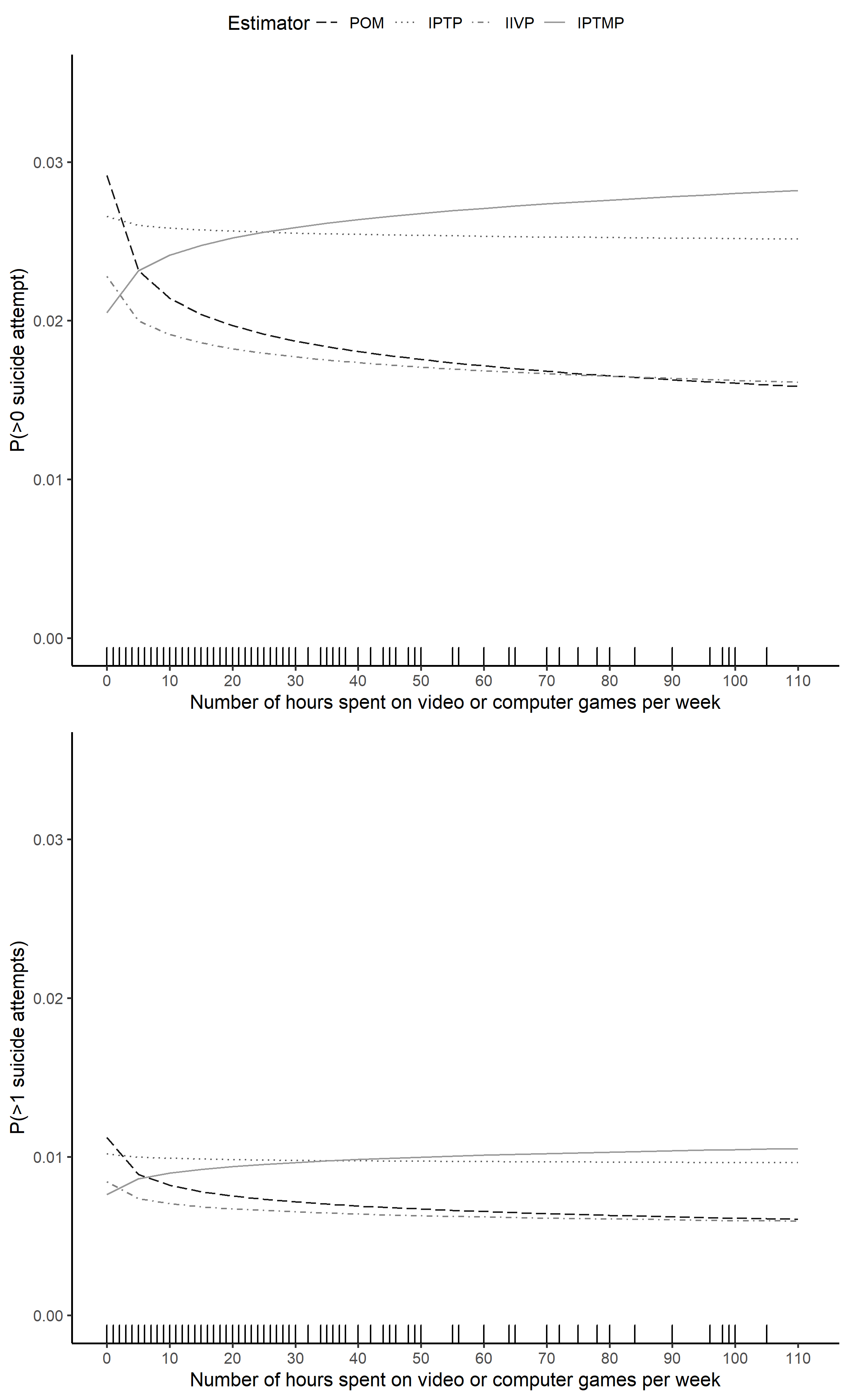}  
   \end{center}
   \caption{Probability of 1 or more suicide attempts (top panel) or of 2 or more suicide attempts (bottom) according to the number of hours spent on video games per week. Comparison of four estimators for the marginal log-OR in the main analysis. The rug plot on the X-axis shows the different values of the number of hours spent playing video games in the study cohort, up to 110 hours per week.} \label{figgh}
  \end{figure}

\section{Discussion}

In this work, we proposed a methodology for causal inference on the marginal effect of a continuous exposure on a categorical, ordinal outcome. Our methodology applies to settings in which monitoring times are not fixed across individuals (i.e., can occur at anytime, for anyone) and in which confounding and selection due to covariate-driven monitoring times can affect the estimator for the exposure effect. We illustrated our methodology via an analysis that aimed to evaluate the causal effect of the time spent on video games, on the number of suicide attempts, in a cohort of individuals followed from adolescence to adulthood. %Although these data are not those from electronic health records, 
Our approach required relatively weak assumptions on the monitoring model to allow inference in settings with both fixed observation times such as in our motivating example as well as irregular observation times, such as would be present in electronic health records data. We found a (non-significant) detrimental effect of a larger time spent on video games on the odds of suicide attempts in the study population.

The methodology we proposed relies on important assumptions about the exposure and the monitoring model. First, although this was not discussed in depth, our estimator relies on the assumption of temporality, where the exposure, to cause the outcome, must have occurred before. In the longitudinal survey, the exposure related to the previous week, and the outcome, to the previous year. Some extrapolation is necessary to assume that the current weekly exposure to video games was relatively similar to that one year ago (for a given individual), and that the effect of interest is well-defined. Secondly, for translating our parametric inference to the causal framework,   we made the assumption of conditional exchangeability for the potential outcome across the different exposure levels, conditional on a certain set of confounders. As in all observational studies, it is possible that unmeasured confounders could not be captured. For instance, other social determinants than those we chose in our models could affect the exposure to video games and the suicide attempts. As discussed in \cite{houtsma2017longitudinal}, features such as previous aggressions, exposure to media violence, fearlessness to pain or about death are risk factors for suicide. These factors could possibly relate to the exposure to video games, either by confounding the effect or by being along the causal path from the exposure to the suicide outcome. Other behavioral characteristics, such as those influenced by the parents during adolescence, were not measured either and could influence both the exposure to gaming and the mood of the individuals. Furthermore, the proposed methodology relies on the normality assumption for the exposure distribution, and therefore, the generalized IPT weight could not adjust fully for confounding if the distribution is far from being normally distributed (even in the cases where we captured all potential confounders). If the distribution is very skewed or multimodals, then other generalized IPT weights such as those assessed or proposed in \cite{naimi2014constructing}, and \cite{schulz2020doubly} should perform better. However, we are confident that our results were not unduly influenced by our assumptions regarding the distribution of the exposure as our sensitivity analyses all came to the same conclusions. We also assumed positivity of the exposure, hypothesizing that all individuals could be exposed to any level of time spent on video games after conditioning on their characteristics. It is unlikely that everyone had the chance of being exposed to more than e.g.~30 to 40 hours of video game per week; this could possibly depend on unmeasured characteristics such as the parents' house rules  or other personal individuals' characteristics (for instance, having a day job). Some methods have been proposed for causal inference on a continuous exposure that relax the strong assumptions on the exposure positivity \citep{haneuse2013estimation,munoz2012population}. 

The study context is also one in which interference may be possible, while we assumed that it was not present. For instance, the exposure of an individual to video games can certainly affect their friend's exposure and, possibly, their friend's outcomes. This remains to study. However, given the sample size relative to the United States population, it is quite unlikely that the participants in the \textit{Add Health} study knew or influenced one another. %For the monitoring model, we made the strong assumption of positivity of monitoring. It is unclear that most studies meet this assumption, especially for those using electronic health records data.

This work is the first to propose a methodology for causal inference data subject to informative monitoring times and confounding when the exposure is continuous, and the outcome, ordinal. Several study outcomes are ordinal, such that our proposed methodology can be useful in several settings. It is also the first substantive study to look at the effect of the time spent weekly on video games on the suicide attempts,  that accounts both for confounding and covariate-driven monitoring times. While other studies have considered potential confounders (without accounting for monitoring times), most presented only a modest adjustment for confounders (see e.g.~\cite{messias2011sadness}). In studies of video game exposure, having informative monitoring times is plausible (see e.g.~\cite{khazaal2014does} who discuss self-selection for online surveys on video games); furthermore, it is very likely that patients with more suicidal ideation may present with different monitoring patterns than others \citep{tylee1999depression}. Thus, it is critical to think about those possible biases in observational studies, when looking at similar causes and their effect on suicide attempts.

\subsection{Acknowledgements} 

This research was enabled in part by support provided by Compute Canada (www.computecanada.ca). Our work is supported by a doctoral scholarship from the Natural Sciences and Engineering Research Council (NSERC) of Canada (Ref. 401223940) to author JC. EEMM acknowledges support from a Discovery Grant from NSERC and a chercheur-boursier career award from the Fonds de recherche du Québec--Santé. RWP acknowledges support from a Discovery Grant from NSERC and a Foundation Scheme Grant from CIHR. 

This research uses data from the \textit{Add Health} program. These data are available in the Data Sharing for Demographic Research repository, at  https://doi.org/10.3886/ICPSR21600.v21. The analysis was restricted to the \textit{Add Health} public-use data. 

\textit{Add Health} was designed by J. Richard Udry, Peter S. Bearman, and Kathleen Mullan Harris at the University of North Carolina at Chapel Hill. The project was funded by the Eunice Kennedy Shriver National Institute of Child Health and Human Development from 1994-2021, with cooperative funding from 23 other federal agencies and foundations. \textit{Add Health} is currently directed by Robert A. Hummer; it was previously directed by Kathleen Mullan Harris (2004-2021) and J. Richard Udry (1994-2004). More information on obtaining \textit{Add Health} data is available on the project website (https://addhealth.cpc.unc.edu). The \textit{Add Health} Parent Study/Parents (2015-2017) data collection was funded by a grant from the National Institute on Aging (RO1AG042794) to Duke University, V. Joseph Hotz (PI) and the Carolina Population Center at the University of North Carolina at Chapel Hill, Kathleen Mullan Harris (PI). The content of this manuscript is solely the responsibility of the authors and does not necessarily represent the official views of the National Institutes of Health or the University of North Carolina at Chapel Hill. 

\newpage

\bibliographystyle{apalike}  
\bibliography{bibiloo.bib}

\section*{Appendix A}
\textbf{Imputation models for missing covariates}\vspace{0.1cm}\\
In the application to the \textit{Add Health} study, covariates were imputed using multivariate imputations by chained equations (MICE), using fully conditional specification, and starting with the covariates with the least missing values, to the most missing values in the condition in the imputation models. The rates of missing values, calculated as the number of missing values in the vector, divided by the length of each covariate vector (for all 4 waves stacked together) are: Sex: 0.0\%; Age: 0.1\%; Race: 0.3\%; ENG: 1.6\%; MATH: 3.6\%; HSS: 5.4\%; GS: 6.2\%; LGR: 18.1\%; RSBI: 18.1\% Feeling depressed: 18.1\%; Video games: 18.9\%; SES: 24.6\%; Smoking: 43.2\%;  HOF: 56.4\%; FHTR: 56.5 \%; FCA 56.7\%. To assess the validity of the imputation and model fits, we used strip plots and density plots. The imputation models provided a good fit, with similar distributions between the observed covariates and the fitted covariates distributions. The strip plots and density plots looked very similar between the observed and fitted distributions for all covariates (except for the variable sex, which fitted distribution was based on only 4 missing values, and for the variable age which presented with slight deviations between the observed and fitted distributions).  
  
  \appendix

\newpage

\noindent
\section*{Supplementary Material A}
\noindent \textbf{Results from the sensitivity analyses: Comparison of the marginal effect of the time spent playing video games on the number of suicide attempts }\vspace{0.2cm}\\  

%Note Janie: This is the analysis ith classical adjustment on X and no log X
\begin{table}[H]
 \begin{center}
\caption{ \textbf{Analysis S1.} Comparison of four estimators for the marginal OR for 1-hour or 10-hour increases in the time spent playing video games per week, on the odds of suicide attempts (number of attempts categorized in 0, 1, or more), \textit{Add Health} study, United States, 1994-2008, $n=6504$. }
\begin{tabular}{ l c  c}
 \hline  \hline
Estimator & 1-hour increase OR (Bootstrap 95\% CI)  & 10-hour increase OR (Bootstrap 95\% CI)  \\ \hline
 $\hat{\beta}_{POM}$ &0.99 (0.97, 1.01)&0.93 (0.70, 1.06)\\
$\hat{\beta}_{IPTP}$ &1.00 (0.97, 1.01)& 1.01 (0.77, 1.15)\\
$\hat{\beta}_{IIVP}$ & 1.00 (0.97, 1.01)& 0.97 (0.75, 1.09)\\
$\hat{\beta}_{IPTMP}$ &1.01 (0.99, 1.02)& 1.11 (0.88, 1.26)\\
    \hline
\end{tabular}
 \label{tab2}
 \end{center}
 \end{table}

\begin{table}[H]
 \begin{center}
\caption{ \textbf{Analysis S2.} Comparison of four estimators for the marginal OR for 1-hour or 10-hour increases in the time spent playing video games per week, on the odds of suicide attempts (number of attempts categorized in 0, 1, or more), \textit{Add Health} study, United States, 1994-2008, $n=6504$.  }
\begin{tabular}{ l c  c}
 \hline  \hline
Estimator & 1-hour increase OR (Bootstrap 95\% CI)  & 10-hour increase OR (Bootstrap 95\% CI)  \\ \hline
 $\hat{\beta}_{POM}$ & 0.99 (0.97, 1.01)& 0.93 (0.70, 1.07)\\
$\hat{\beta}_{IPTP}$ &1.01 (0.98, 1.02)& 1.07 (0.84, 1.20) \\
$\hat{\beta}_{IIVP}$ &1.00 (0.97, 1.01) & 0.97 (0.76, 1.09)\\
$\hat{\beta}_{IPTMP}$& 1.01 (0.99, 1.02)&1.08 (0.87, 1.20) \\
    \hline
\end{tabular}
 \label{tab2}
 \end{center}
 \end{table}
 
 \begin{table}[H]
 \begin{center}
\caption{\textbf{Analysis S3.} Comparison of four estimators for the marginal OR for a two-fold or 8-fold increases in the time spent playing video games per week, on the odds of suicide attempts   (number of attempts categorized in 0, 1, or more), \textit{Add Health} study, United States, 1994-2008, $n=6504$. }
\begin{tabular}{ l c  c}
 \hline  \hline
Estimator & 2-fold OR (Bootstrap 95\% CI)  & 8-fold OR (Bootstrap 95\% CI)  \\ \hline
 $\hat{\beta}_{POM}$ & 0.91 (0.82, 0.99)& 0.76 (0.55, 0.98)\\
$\hat{\beta}_{IPTP}$ & 1.00 (0.88, 1.09)& 0.99 (0.69, 1.30)\\
$\hat{\beta}_{IIVP}$ & 0.95 (0.85, 1.04)& 0.86 (0.61, 1.11)\\
$\hat{\beta}_{IPTMP}$ &1.03 (0.91, 1.13)&1.09 (0.74, 1.44) \\
    \hline
\end{tabular}
 \label{tab2}
 \end{center}
 \end{table}
\newpage

\section*{Supplementary Material B}
\noindent \textbf{Results from the main analysis and the sensitivity analyses: Graphs of the estimated probability of 1 or more, or of 2 or more suicide attempts, by the number of hours spent playing video games weekly} \vspace{0.2cm}\\
  \newpage
\noindent   \textbf{Main analysis with the log-OR estimated for a 1-unit increase in $\log_2$ (number of hours spent playing video games$+1$) and the use of a generalized inverse probability of treatment weight for modelling $\log_2$ (number of hours spent playing video games$+1$) as a continuous variable, along with  the 95\% confidence interval bands added}
  
  \begin{figure}[H]
   \begin{center}
  \includegraphics[width=9cm]{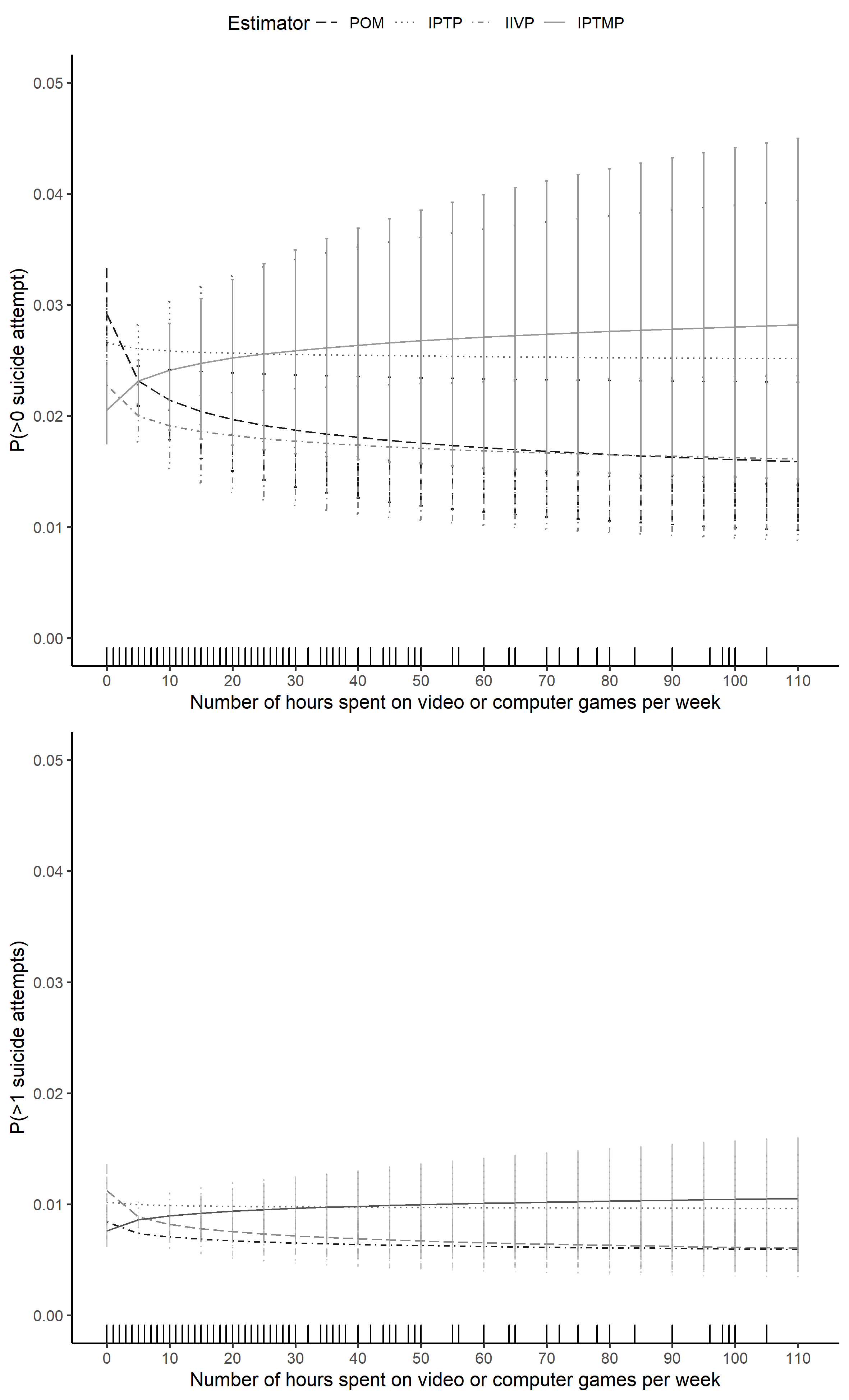}  
   \end{center}
   \caption{\textbf{Main analysis.} Probability of 1 or more suicide attempts (top panel) or of 2 or more suicide attempts (bottom) according to the number of hours spent playing video games per week. Comparison of four estimators for the marginal log-OR. The bands around the point estimates correspond to 95\% CIs computed using the bootstrap percentiles. The rug plot on the X-axis shows the different values of the number of hours spent playing video games in the study cohort, up to 110 hours per week.} \label{figgh}
  \end{figure}
  \newpage
 
  \noindent  \textbf{Sensitivity analysis 1 (S1): The number of hours spent playing video games as a continuous exposure incorporated in the outcome model, and the use of a generalized inverse probability of treatment weight for the continuous exposure (under the normality assumption for the number of hours spent playing video games weekly) } 
  
  \begin{figure}[H]
   \begin{center}
  \includegraphics[width=10cm]{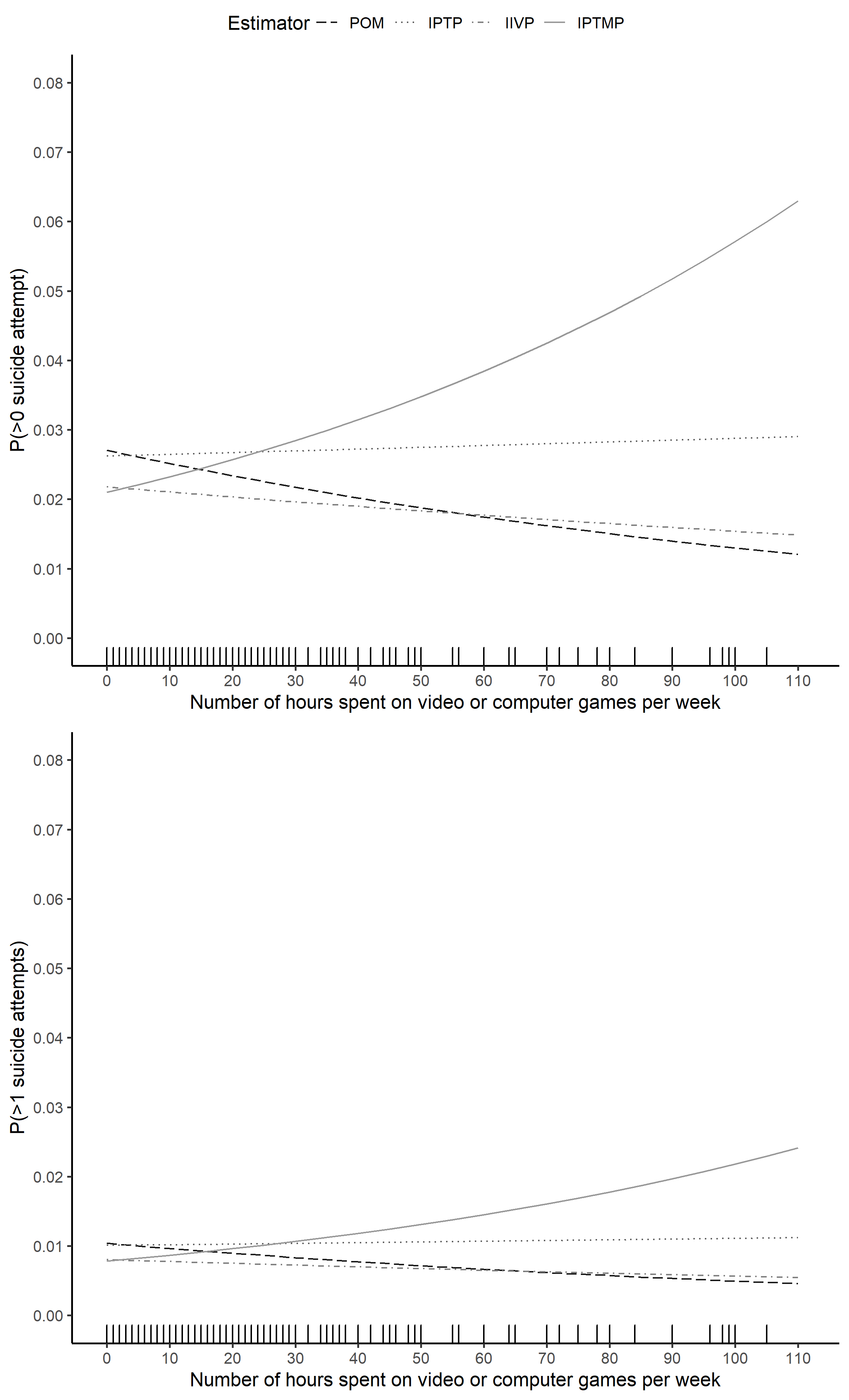}  
   \end{center}
   \caption{\textbf{Analysis S1.} Probability of 1 or more suicide attempts (top panel) or of 2 or more suicide attempts (bottom) according to the number of hours spent playing video games per week. Comparison of four estimators for the marginal log-OR. %The bands around the point estimates correspond to 95\% CIs computed using the bootstrap percentiles. % and re-transforming using the expit function (which accounts for the variance of all coefficients in the prediction). 
   The rug plot on the X-axis shows the different values of the number of hours spent playing video games in the study cohort, up to 110 hours per week.} \label{figgh}
  \end{figure}
  
  \begin{figure}[H]
   \begin{center}
  \includegraphics[width=10cm]{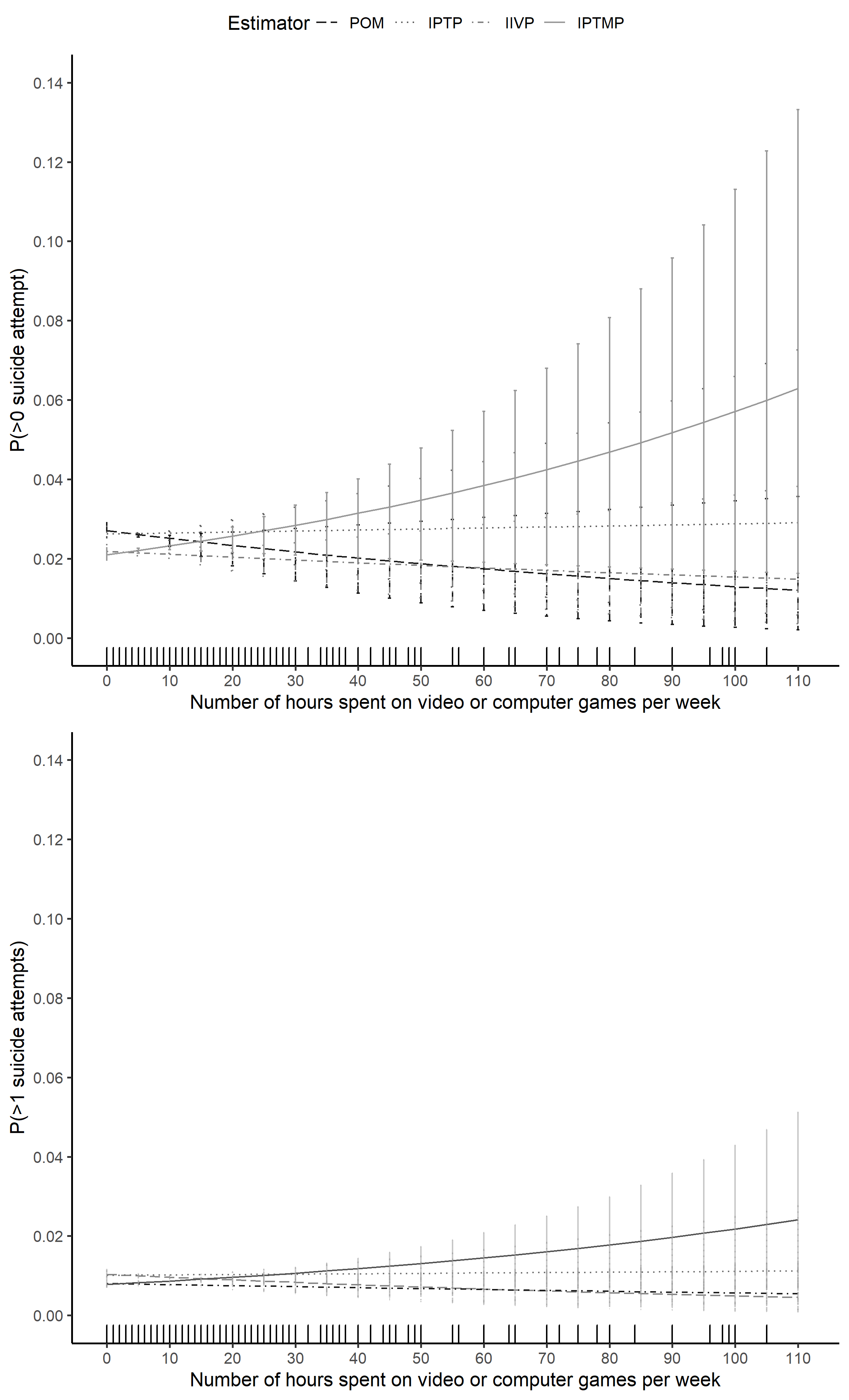}  
   \end{center}
   \caption{\textbf{Analysis S1.} Probability of 1 or more suicide attempts (top panel) or of 2 or more suicide attempts (bottom) according to the number of hours spent playing video games per week. Comparison of four estimators for the marginal log-OR. The bands around the point estimates correspond to 95\% CIs computed using the bootstrap percentiles. % and re-transforming using the expit function (which accounts for the variance of all coefficients in the prediction). 
   The rug plot on the X-axis shows the different values of the number of hours spent playing video games in the study cohort, up to 110 hours per week.} \label{figgh}
  \end{figure}
  \newpage
\noindent   \textbf{Sensitivity analysis 2 (S2): The number of hours spent playing video games as a continuous exposure incorporated in the outcome model, and the use of a generalized inverse probability of treatment weight for the categorical exposure (five categories)} \\

  \begin{figure}[H]
   \begin{center}
  \includegraphics[width=10cm]{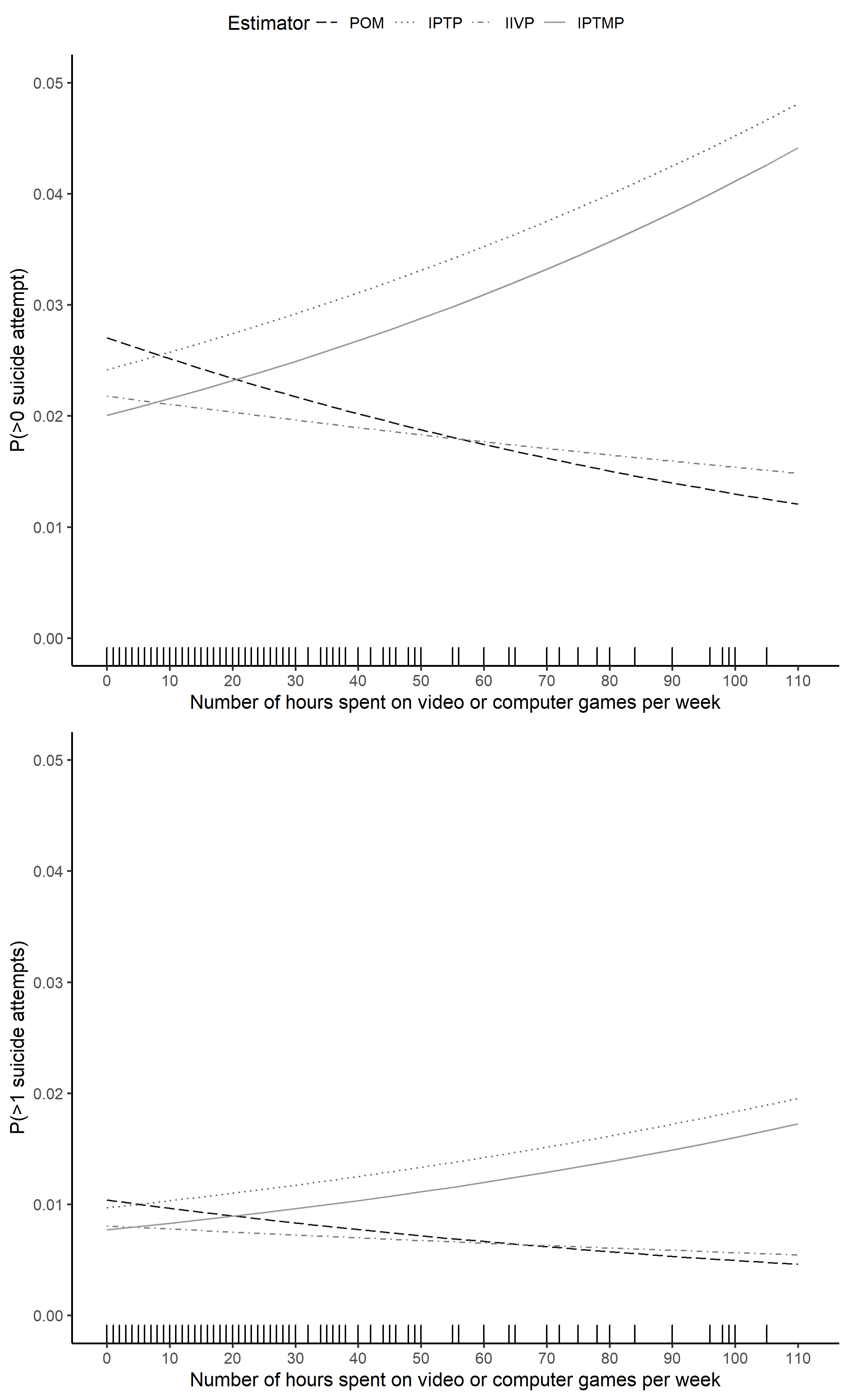}  
   \end{center}
   \caption{\textbf{Analysis S2.} Probability of 1 or more suicide attempts (top panel) or of 2 or more suicide attempts (bottom) according to the number of hours spent playing video games per week. Comparison of four estimators for the marginal log-OR. %The bands around the point estimates correspond to 95\% CIs computed using the bootstrap percentiles. % and re-transforming using the expit function (which accounts for the variance of all coefficients in the prediction). 
   The rug plot on the X-axis shows the different values of the number of hours spent playing video games in the study cohort, up to 110 hours per week.} \label{figgh}
  \end{figure}
  
  \begin{figure}[H]
   \begin{center}
  \includegraphics[width=10cm]{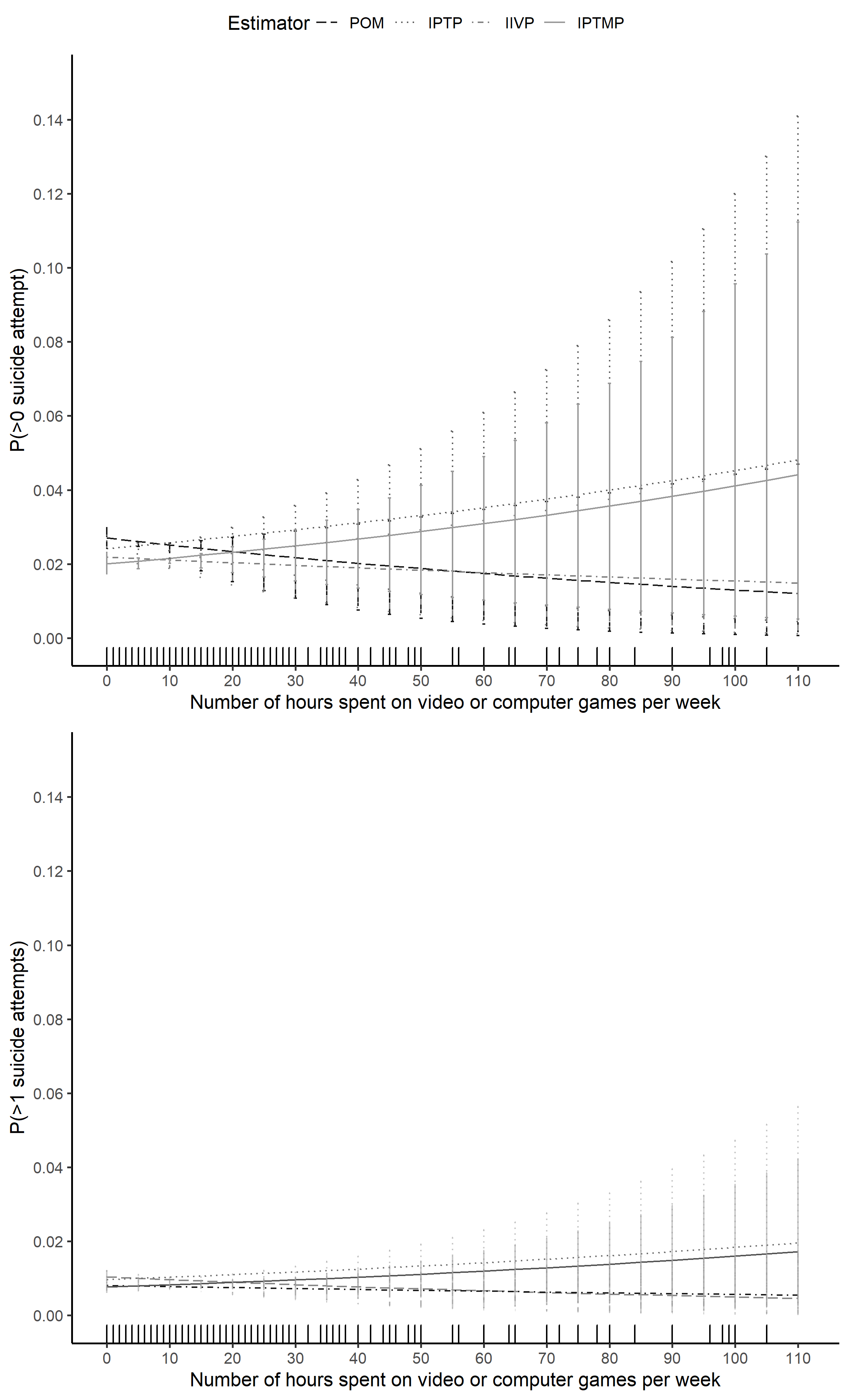}  
   \end{center}
   \caption{\textbf{Analysis S2.}  Probability of 1 or more suicide attempts (top panel) or of 2 or more suicide attempts (bottom) according to the number of hours spent playing video games per week. Comparison of four estimators for the marginal log-OR. The bands around the point estimates correspond to 95\% CIs computed using the bootstrap percentiles. % and re-transforming using the expit function (which accounts for the variance of all coefficients in the prediction). 
   The rug plot on the X-axis shows the different values of the number of hours spent playing video games in the study cohort, up to 110 hours per week.} \label{figgh}
  \end{figure}
  \newpage
  \noindent   \textbf{Sensitivity analysis 3 (S3):  The $\log_2$(number of hours spent playing video games$+1$) as a continuous exposure incorporated in the outcome model, and the use of a generalized inverse probability of treatment weight for the categorical exposure (five categories)}
    \begin{figure}[H]
   \begin{center}
  \includegraphics[width=10cm]{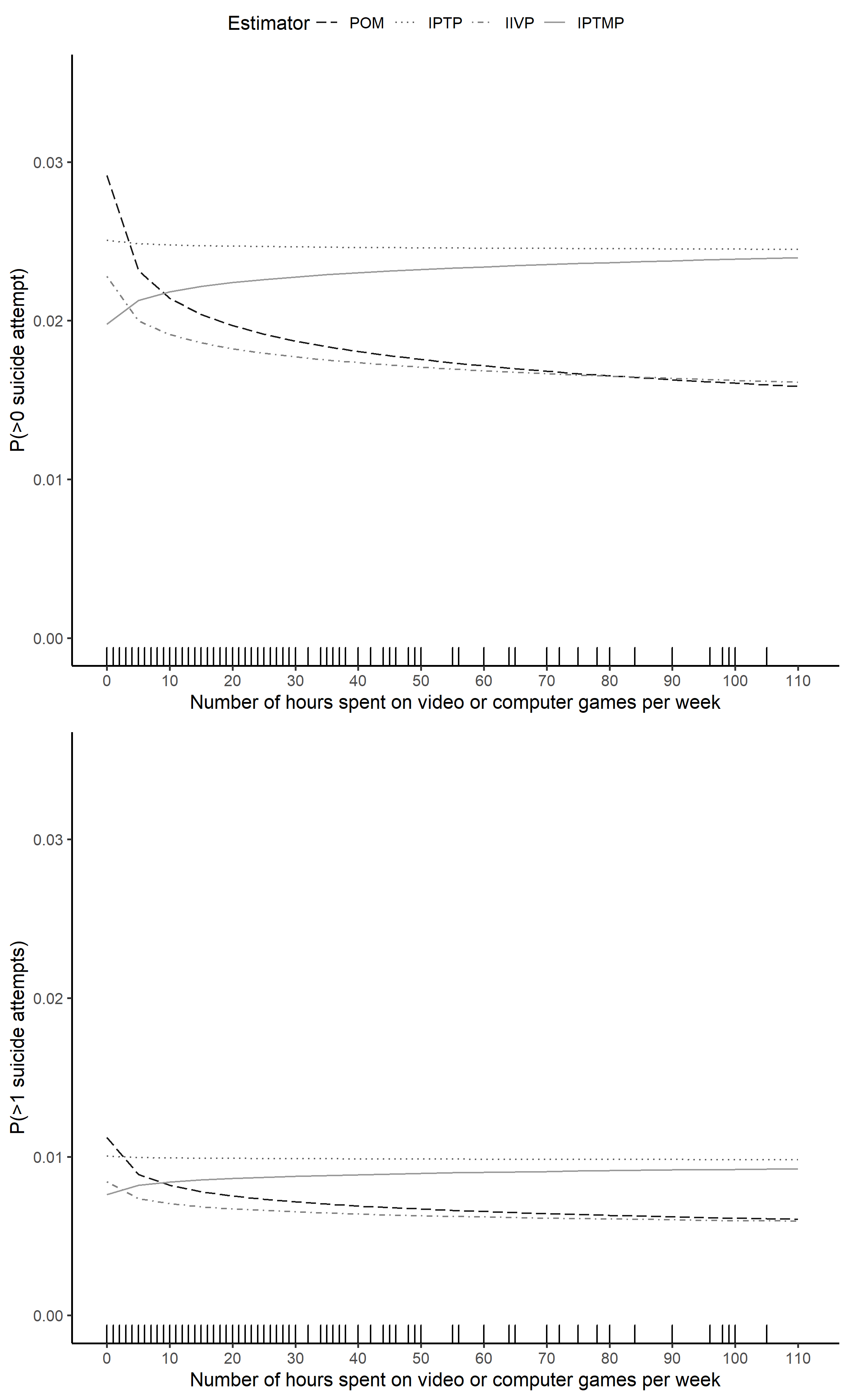}  
   \end{center}
   \caption{\textbf{Analysis S3.} Probability of 1 or more suicide attempts (top panel) or of 2 or more suicide attempts (bottom) according to the number of hours spent playing video games per week. Comparison of four estimators for the marginal log-OR. %The bands around the point estimates correspond to 95\% CIs computed using the bootstrap percentiles. % and re-transforming using the expit function (which accounts for the variance of all coefficients in the prediction). 
   The rug plot on the X-axis shows the different values of the number of hours spent playing video games in the study cohort, up to 110 hours per week.} \label{figgh}
  \end{figure}
  
  \begin{figure}[H]
   \begin{center}
  \includegraphics[width=10cm]{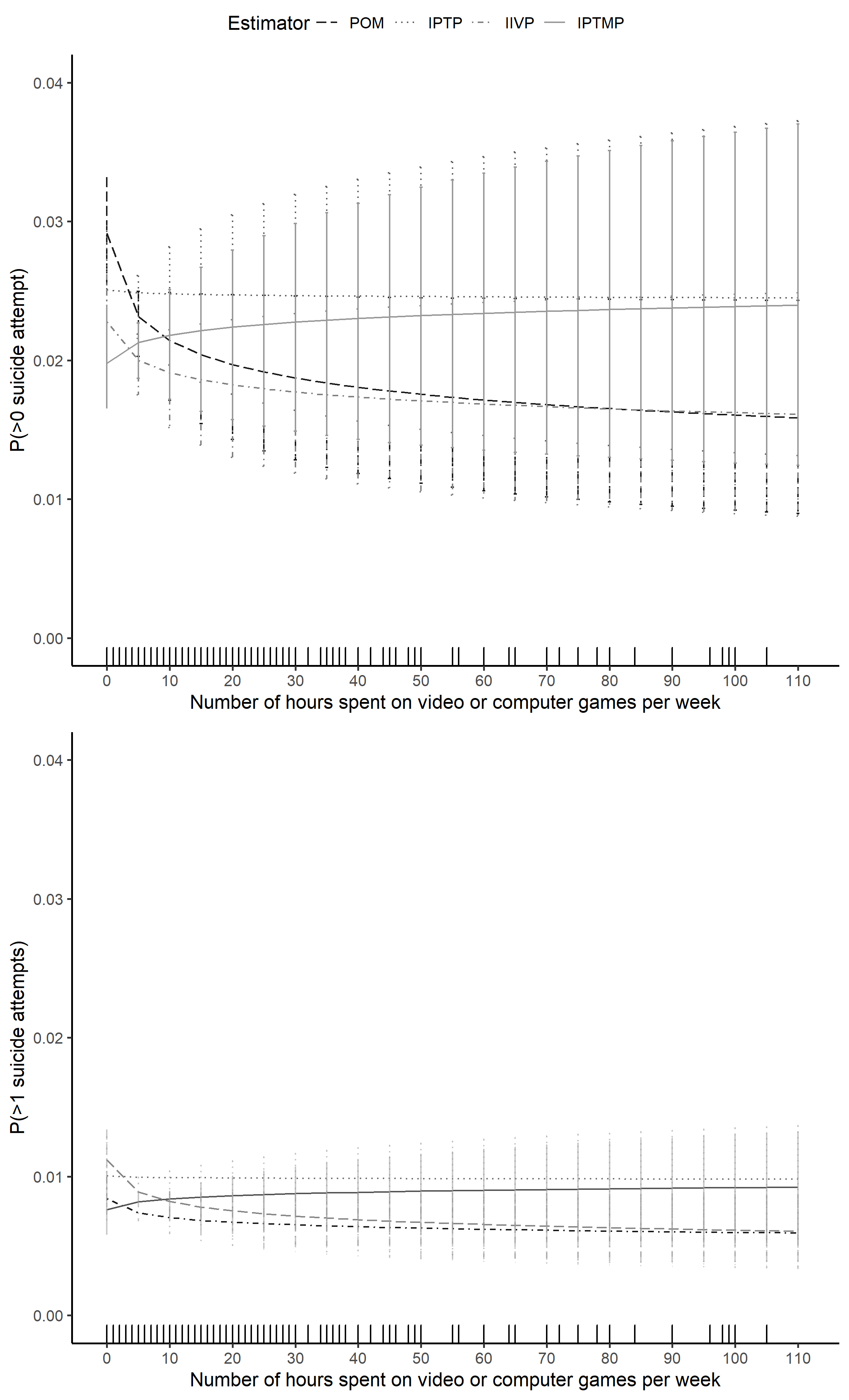}  
   \end{center}
   \caption{\textbf{Analysis S3.} Probability of 1 or more suicide attempts (top panel) or of 2 or more suicide attempts (bottom) according to the number of hours spent playing video games per week. Comparison of four estimators for the marginal log-OR. The bands around the point estimates correspond to 95\% CIs computed using the bootstrap percentiles. % and re-transforming using the expit function (which accounts for the variance of all coefficients in the prediction). 
   The rug plot on the X-axis shows the different values of the number of hours spent playing video games in the study cohort, up to 110 hours per week.} \label{figgh}
  \end{figure}
 
\end{document}